\documentstyle[12pt,aaspp4]{article}

\received{}
\accepted{}
\slugcomment{Submitted to The Astrophysical Journal}

\begin{document}
\include{psfig}
\def\kms{km~s$^{-1}$ }
\def\Lya{Ly$\alpha$ }
\def\lya{Ly$\alpha$}
\def\Lyb{Ly$\beta$ }
\def\lyb{Ly$\beta$}
\def\Lyg{Ly$\gamma$ }
\def\Lyd{Ly$\delta$ }
\def\Lye{Ly$\epsilon$ }
\def\d{$d_5$ }
\def\Ly{Lyman}
\def\ang{\AA }
\def\gq{$\geq$ }
\def\zem{$z_{em}$ }
\def\zabs{$z_{abs}$ }
\def\cm2{cm$^{-2}$ }
\def\hi{\ion{H}{1} }
\def\di{\ion{D}{1} }
\def\nht{N(\ion{H}{1})$_{total}$ }
\def\nhi{N(\ion{H}{1})}
\def\cmin{$\chi^2_{min}$ }
\def\g3{$>3.0$}
\def\etal{et al. }

\title{The Deuterium Abundance towards QSO 1009+2956}

\author{Scott Burles\altaffilmark{1} \& David Tytler\altaffilmark{1}}
\affil{Department of Physics, and Center for Astrophysics and Space
Sciences \\
University of California, San
Diego \\
C0111, La Jolla, California, 92093-0111}

\altaffiltext{1}{Visiting Astronomer, W. M. Keck Telescope, California
Association for Research in Astronomy.}

%
% -- 2.Abstract page!
%

\begin{abstract}
We present a measurement of the deuterium
to hydrogen ratio (D/H) in a metal-poor absorption system 
at redshift $z=2.504$ towards the QSO 1009+2956.
We apply the new method of Burles \& Tytler (1997)
to robustly determine D/H in 
high resolution \Lya forest spectra, and include a constraint on the
neutral hydrogen column density determined from 
the Lyman continuum optical depth in low resolution spectra.
We introduce six separate models to measure D/H and to 
assess the systematic dependence on 
the assumed underlying parameters.  We find that the deuterium
absorption feature contains a small amount of contamination from
unrelated \ion{H}{1}.  Including the effects of the contamination, we calculate 
the 67\% confidence interval of D/H in this absorption system,
log (D/H) $= -4.40 ^{+0.06}_{-0.08}$.
This measurement agrees with the low measurement by Burles \& Tytler (1997)
towards Q1937--1009, and the combined value gives the best determination
of primordial D/H, log (D/H)$_p = -4.47 ^{+0.030}_{-0.035}$
or D/H $= 3.39 \pm 0.25 \times 10^{-5}$.
Predictions from standard big bang nucleosynthesis (SBBN)
give the cosmological 
baryon to photon ratio, 
$\eta = 5.1 \pm 0.3 \times 10^{-10}$, and
the baryon density in units of the critical density, 
$\Omega_b\,h^2 = 0.019 \pm 0.001$, where $H_0 = 100\,h$ \kms Mpc$^{-1}$.
The measured value of
(D/H)$_p$ implies that the primordial abundances of both
$^4$He and $^7$Li are high, and consistent with some recent studies. 
Our two low measurements of primordial D/H 
also place strong constraints on inhomogeneous models of big bang
nucleosynthesis.
\end{abstract}

\section{INTRODUCTION}

Measurements of the deuterium-to-hydrogen ratio (D/H) in QSO absorption
systems can test current theories of the early universe, in
particular, the theory of big bang nucleosynthesis (BBN, c.f.
\cite{ful96}; \cite{sch97} and references therein). 
The amount of deuterium produced during the epoch of
BBN is very sensitive to the
to the cosmological baryon-to-photon ratio, $\eta$ (\cite{ree73};
\cite{eps76}).  A sample of D/H measurements in QSO absorption systems
can establish the primordial abundance ratio of D/H and constrain 
non-standard models of BBN with limits on 
the spatial variation of $\eta$ (\cite{jed94b}, \cite{jed95}).    
A well constrained primordial D/H value places tight constraints
on $\eta$ and gives
predicted primordial abundances of the other
light elements, which can be compared with current 
observations to test the paradigm of SBBN (\cite{card96}; \cite{hat97}).

With high-resolution spectroscopy of distant QSOs, we can measure
the deuterium-to-hydrogen ratio (D/H) in select absorption systems
which lie along the line of sight (\cite{ada76}).  
High redshift metal-poor QSO absorption systems are ideal sites to infer
a primordial value for D/H (\cite{jed97b}).
To detect absorption in D-\Lya, the absorption
system must 
have a large hydrogen column density \nhi $> 10^{17}$ \cm2 and
an usually narrow velocity structure (c.f. Tytler \& Burles 1997,
hereafter TB).
We discovered such an absorption system at $z=2.504$ towards
the bright QSO 1009+2956 ($z_{em}$ = 2.63, V=16.0).
TB performed a preliminary analysis and measurement of D/H in this system.
A single consistent absorption model accounted for the H and D absorption
in the strongest Lyman lines (\Lya -- \Lyg), the metal line profiles
of C and Si, and the Lyman continuum break arising from the large \hi column.
To constrain the hydrogen velocity structure, TB used a two component
model for H and D with velocity positions given by the metal line components.
TB found log (D/H) = $-4.52 \pm 0.08 \pm 0.06$
(1$\sigma$ statistical and systematic errors).
TB also found that the profile
of the deuterium feature suggested contamination from \hi, and estimated
that D/H is likely lower by 0.08 dex.

Studies of other QSO absorption systems at high redshift give
limits on D/H which are consistent with our measurements.
Songaila \etal (1994) and Carswell \etal (1994) reported a
detection of deuterium in the absorption system 
at $z=3.32$ towards Q0014+8118.
Rugers \& Hogan (1996a) reanalyzed the system and measured
D/H in two separate components and found D/H = $19 \pm 0.5 \times 10^{-5}$
in both components.  We reobserved Q0014+8118 and found the expected
position of D was 
contaminated with \hi absorption, 
and only an upper limit of D/H $< 30 \times 10^{-5}$ 
could be extracted from this system (Tytler \etal 1997).
Another system at $z=2.80$ towards Q0014+8118 was analyzed by Rugers
\& Hogan (1996b).  The \Lya absorption feature is the only Lyman line 
accessible, and we showed that this complex system
cannot give any practical constraints on D/H (Tytler \etal 1997).
Wampler \etal (1996) find $D/H < 15 \times 10^{-5}$
at $z=4.672$ towards  
Q1202$-$0725.
Carswell \etal (1996) find a lower
limit of D/H $> 2 \times 10^{-5}$ at $z=3.086$ towards
Q0420$-$3851.  
Webb et al. (1997) recently deduced a D/H value at $z=0.701$
towards the low redshift QSO
1718+4807 using a spectrum obtained with the Hubble Space Telescope
(HST).
Unlike Q1937--1009 and Q1009+2656, only one \hi line was observed, so
the velocity structure of the H is not well known.
Assuming a single component fit to the \Lya , they find
D/H = $20 \pm 5 \times 10^{-5}$, but this value will remain suggestive
unless confirmed with approved HST observations of the high-order Lyman
lines.  All the above limits on D/H are consistent with 
our measurements.  

The present analysis supersedes that of TB for the following 
reasons.  We have obtained additional spectral data of Q1009+2956,
including full coverage of the Lyman series and better quality data
above and below the Lyman continuum break. 
We consider six new absorption models, 
each with different underlying assumptions about the number of
velocity components, free parameters, and exterior constraints.
We use the new method to measure D/H in QSO absorption systems
described in Burles \& Tytler (1997a, hereafter BT).
In Sec. 2, we describe the observations and reductions of the spectral data.
In Sec. 3, we present the detailed analysis of this absorption system, and
discuss the results of the D/H measurement with each of the six new models.
in Sec. 4, we combine these results with another measurement of D/H towards
Q1937--1009 (\cite{tyt96}) to give the tightest constraints on the primordial
value of D/H, and calculated the predicted values of $\eta$,
$\Omega_b$, and the other primordial light element abundances with calculations
from SBBN. In Sec. 5, we discuss the implications of low detuerium
measurements in two QSO absorption systems and constraints on 
inhomogeneities during BBN.
 
\section{OBSERVATIONS AND DATA REDUCTION}

In this analysis, we use two independent data sets to analyze
the absorption system at $z=2.504$ towards
Q1009+2956 (\zem=2.63, V=16.0).
The first data set is a high signal-to-noise ratio (SNR),
high-resolution spectrum (FWHM = 8 \kms) obtained 
with HIRES at the W. M. Keck 1 10-m Telescope.
The high-resolution spectrum fully resolves each absorption
line in the Lyman series at $z=2.504$ and allows a detailed analysis
of the velocity structure of the absorber, which is required to measure
D/H accurately.
Six exposures with HIRES, totaling 12.4 hours, are outlined in Table 1.
The processing of the CCD images and the extraction of spectra is described in
Tytler \etal (1996).  
Each spectra was normalized to an initial estimate of the quasar continuum
by fitting a Legendre polynomial to the unabsorbed regions of the spectra.
The spectra were added together, using weights proportional to the
SNR to produce a final spectrum.

The second data set is a high SNR, low-resolution spectrum
(FWHM = 4 \AA) of Q1009+2956,
which has been calibrated to an absolute flux scale and has increased
sensitivity below 3200 \AA, obtained with the Kast Double Spectrograph on 
the Shane 3-m Telescope at Lick Observatory.  We took
a total of five hours of data in six separate exposures on the nights
of November 28, 1995, and December 15, 1996.  The last four exposures
used a long (145 arcsec) and wide (9 arcsec) slit
to ensure that slit losses were negligible, which 
gave absolute flux calibration with multiple standard
star spectra obtained with the same setup on the same nights.
The other spectra were taken with a long narrow (2 arcsec)
slit, and were corrected
to match the overall flux calibration and smoothed to match the
spectral resolution of the wide-slit spectra.
The spectra were optimally extracted from the two-dimensional CCD images, 
after the images were baseline and bias subtracted,
and then flat-field corrected.  The spectra were set onto a vacuum, 
heliocentric wavelength scale using images of Ne-Ar and He-Hg-Cd Arc lamps.
The final spectrum is shown in Figure 1, and will be referred to as
the Lick spectrum throughout this paper.  

\section{THE MEASUREMENT OF D/H}

In this section, we outline the method to measure D/H in the absorption
system at $z=2.504$ towards Q1009+2956.  
A more detailed description of the general method can be found in BT.
First, we measure the total hydrogen column density, \nht, in the
D/H absorption system (DHAS).  \nht~ represents the sum of the \hi column
densities in all of the absorbing components in this system.
We obtain D/H from models of the Lyman series absorption
lines in the HIRES spectrum, which use \nht ~as a constraint.

\subsection{Total Hydrogen Column Density}

We use the Lick spectrum 
and the method described in Burles \& Tytler (1997b)
to measure the Lyman continuum optical depth of the DHAS towards Q1009+2956.
To measure
the Lyman continuum optical depth, we need to determine the QSO continuum
above and below the Lyman break at $912 (1 + z_{abs})$ \AA.
We divide the Lick spectrum by the HIRES spectrum to 
give the unabsorbed continuum level above the break 
(Burles \& Tytler 1997b).  

Figure 2 shows the \Lya forest in the Lick spectrum
(top panel) and the HIRES spectrum smoothed to the spectral resolution
of the Lick spectrum (bottom panel).  The crosses in the top panel show
the pixels of the Lick spectrum after dividing by the normalized flux
level in the smooth HIRES spectrum, and they represent the intrinsic
unabsorbed QSO continuum of Q1009+2956.  

This continuum of Q1009+2956 shows more structure than Q1937--1009 studied by
Burles \& Tytler (1997b).  The features seen in Figure 2 
are likely intrinsic QSO emission lines (Lyman series, \ion{Fe}{3}, \ion{C}{3}).
By removing the intervening absorption features,
we can discover and measure new emission lines in high redshift QSO spectra.
The slope of the underlying power-law continuum shows a 
break near 3700 \AA (1020 \AA rest), which was also seen in Q1937--1009,
and by Zheng et al. (1997) in the spectra of lower redshift QSOs.
We made a linear fit to the continuum blueward
of 3700 \AA, and the best fit is shown as the straight solid line in Figure 2.
We do not use a more complex function because we can not predict
the shape of the  unabsorbed continuum plus emission lines
at $\lambda < 3200$ \AA.
The straight line represents the approximate continuum level in the
spectral region below 3700 \AA.  The standard deviation of the
points about the line is 10\%, and we use this as the uncertainty of the
extrapolated continuum below 3200 \AA.  

We use the extrapolated continuum to normalize the flux below the Lyman
limit, which is shown in Figure 3.  Each data point represents the 
normalized flux in each pixel in the Lick spectrum with 
1$\sigma$ error bars.  
We calculate \nht in the D/H absorption system using a maximum
likelihood method.
The model fit includes
higher order Lyman lines from absorbers measured at higher wavelengths
in the HIRES spectrum, and from a random sampling of the \Lya forest over
the region shown in Figure 3.  
The rest of the Lick spectrum
shown in Fig. 3 is consistent with the model constructed from all absorbers
at higher redshift ($z > 2.00$), and the extra absorption (c.f. at 3155 \AA)
is accounted for with random Lya lines with redshifts between $1.58 < z < 1.64$.
The drop in flux near 3140 \AA is due to a second Lyman limit
system at $z=2.43$.
We find log[\nht] = 17.39 $\pm$ 0.06 \cm2 (1$\sigma$ error including
the 10\% continuum uncertainty).  The dotted lines in Fig. 3 represent the
1$\sigma$ variation in \nht.

We compare the measurement
of \nht~ to an independent determination using the 
the highest
order Lyman lines in the HIRES spectrum.  We apply a simple
two component model to the line profiles of 
Ly-12 to Ly-15. 
The best fit to the HIRES spectrum gives
log \nhi = 17.35 $\pm$ 0.07 \cm2, which is consistent with the Lyman
continuum measurement.

In the following analysis we use \nht as a new constraint on 
models of the absorption system. The \nhi in these
models must fit the individual Lyman series lines, and be
consistent with \nht.

\subsection{Constraining D/H}

We use a $\chi^2$ minimization routine described by 
BT, and six new models of the absorption system. 
As parameters change during the minimization, new model
spectra are calculated and compared to the 
HIRES spectrum in the regions listed in Table 2.
Each model includes two types of absorbers:
\hi lines with enough \nhi to show D and/or 
contribute to the total hydrogen column, \nht~ (``main'' components).
The other type are
extra \hi lines with low \nhi, which do not contribute significantly
to \nht and do not show D.

All absorbers are modeled as Voigt profiles (\cite{spi78}).
The main components are parameterized by column densities 
N(\ion{D}{1}) and N(\ion{H}{1}), temperature, T, turbulent
velocity, $b_{tur}$, and redshift, $z$.
During the fitting procedure, we assume that D/H is equal
in all components, so the column density of D is always given
by N(\ion{D}{1}) = N(\ion{H}{1}) $\times$ (D/H).
We model the Lyman lines without using the metal lines, and 
return to the metal line analysis 
after the model fitting of the Lyman series is finished.
The other \hi absorbers, listed in Table 3,
are described by three free parameters,
column density (N), total velocity dispersion ($b$), and redshift
($z$).  The unabsorbed quasar continuum is described 
by Legendre polynomials
of order n, where n is shown in Table 2 for each region.
The continuum order is higher in regions with more pixels and higher
SNR (i.e. \Lya)
During the initial data reduction, full echelle spectral orders (2048 pixels)
were normalized using fifth order Legendre polynomials.  

Figure 4 shows the HIRES spectrum covering 
the Lyman series lines stacked in velocity space.
By stacking the data in velocity space, the alignment of
the Lyman absorption lines is evident.  
The HIRES spectrum spans the entire Lyman series, but we
have chosen regions which are free from excessive blending
from random \Lya absorption along the line of sight.
Fortunately, we can include the majority of high order Lyman lines,
Ly-12 and above, which place the tightest constraints on the
\hi absorption in the main components.  The Lyman lines become
unsaturated in their line centers at Ly-20 and above.
In Figure 5, we show the HIRES spectrum containing these
high-order Lyman lines. 
A large unassociated system, at $z=1.6355$,
absorbs all the flux in the region
containing the Lyman lines, Ly-16 through Ly-19.

In Figures 4 to 6, the smooth grey line which closely traces the data 
represents the best fit of Model 2 (discussed below).
The entire Lyman series
is well fit by the four main \hi components (black tick marks). 
Three main components lie near zero velocity, which is at $z= 2.503571$,
and the fourth lies redward at $\Delta\, v = 66$ \kms.
Each main \hi component has a
corresponding deuterium absorption line (grey tick marks) at a relative velocity
of $-81.6$ \kms.  The deuterium absorption is significant only in
the regions of \Lya and \Lyb.
Figure 6 shows the separate regions listed in Table 2 and used to constrain
the model fits.  In Figs. 6a and 6b, we also show the individual profiles
of the \hi and \di absorption features.  
We start each minimization with the unabsorbed continuum (normalized to unity)
shown in Figure 6, then we allow the coefficients of the polynomials
to vary during the minimization.

We consider six different models to test the systematic 
dependence and the sensitivity of the D/H measurement on the
underlying assumptions in each model.  Each model is constrained by
the HIRES data in the 
spectral regions in Table 2, and \nht~ given by the optical
depth of the Lyman continuum absorption in the Lick spectrum. 
Model 1 has the fewest absorbers, and is the least
complex.  It includes three main components
and allows for free parameters in the continuum specified by the number
of free parameters listed in Table 2.  The second and third models
are identical to Model 1, but they include four and five main components,
respectively.   
The remaining Models (4--6) are identical to Model 2, with one 
exception each.
Model 4 does not
use the constraint on \nht~ given by our analysis of the Lick spectrum.
By excluding the constraint on \nht~, we can 
test how sensitive our measurement of D/H is to the \nht~ constraint.
Model 5 does not allow the continuum to vary from the initial continuum
estimates.  
Model 6 includes extra \hi absorption to account for possible contamination
of the D-\Lya feature.

For each model, we calculate \cmin for values of D/H from log D/H = $-4.95$
to $-4.00$ in steps of 0.01 dex.   We derive confidence levels
on D/H by the method of $\Delta \, \chi^2$.  Table 4 summarizes
the results of the \cmin calculations for all six models.
We show the values of D/H which fall in the 95\% confidence region,
the minimum $\chi^2$, along with the overall minimum $\chi^2$ for each
model and the number of free parameters in each model.

The major results from our analysis with the six different models
are shown graphically in Fig. 7, with four different
parameters as a function of D/H:  $\chi^2$, N(\ion{H}{1}), $b$(\ion{H}{1}), and
$\Delta \, v$, where $\Delta \, v/c = \Delta \, z / (1+z_0)$ and 
$z_0 = 2.503571$.  
By presenting the parameters as a function of D/H, we determine their
sensitivity to D/H.

Model 1 is very similar to the absorption model analyzed by TB.
In the analysis of TB,
only spectral coverage of \Lya, \Lyb, and \Lyg was available, so 
the positions of the hydrogen components were poorly constrained.
TB assumed that the two main \hi components were matched the velocity
positions of two component metal lines,  at
$\Delta\, v = 0, 11$ \kms in Fig. 7.  The assumption
was justified by the predicted deuterium absorption position, but
systematic errors could have been introduced.
In the present analysis, the positions of the main components are free
parameters, and are constrained only by the absorption profiles of the
Lyman series.  Fig. 7a shows variation
of velocity position as a function of D/H.  The weaker component
(black solid line) does agree with the position of the 
red metal lines near $\Delta\, v = 11$ \kms, 
but the stronger component at $\Delta\, v =6$ \kms does not match 
the blue metal lines.  The third component at $\Delta\, v = 72$ \kms
is required to model the absorption in the red wing of the Lyman  
lines.  This component becomes optically thin at Ly-6 and has well
constrained parameters that are insensitive to the value of D/H.
But the \nhi from the third component contributes to \nht, and requires
that we include this absorber as a main component.  Because its
parameters are fairly insensitive to differences between the six different
models, we do not display its parameters in the remaining five models in
Fig. 7.

Model 2 includes another main \hi component, which falls
between the two \hi components of Model 1 (Fig. 7b),
and contains about 10\% of the total hydrogen column.  
This extra component allows a better overall fit to the HIRES spectrum,
which gives \cmin = 906.1 (Table 4). 
Also, the extra component allows the strongest \hi component to move
near $\Delta\, v = 0$ \kms, but compromises the position of the 
redward component.

In Fig. 6, we show the individual HIRES spectral regions with the 
model spectra giving the best fit of Model 2.  
The model gives a good fit to the data in
all of the regions, and the continuum in the best fit model agrees 
with the initial continuum 
estimate in all regions except Ly-6.  The drop in continuum in the region
containing Ly-6 is likely to an additional absorber on the redward side,
which was not included in any of the models.
The fit is adequate is the Lyman limit region, due to the low SNR in this
region, and additional absorbers are not required.
The model continuum 
is constrained by the pixels showing little or no absorption.  
These pixels are at the edges of the region in \Lya (Fig. 6a), and the
continuum is specified by the optical depth in the damping wings, which
is proportional to \nht.  The shape of the best fit model continuum suggests
that the initial level of continuum was slightly overestimated.   
The \hi and \di absorption profiles are shown as dot-dashed and dashed lines,
respectively, in Figs. 6a and 6b (\Lya and \Lyb).  Both profiles represent
the total absorption of the three strongest components in Model 2
(components shown in Fig. 7b).  Notice
that the deuterium profile is asymmetric, with a steeper blue wing,
like the metal lines. This qualitatively validates the TB assumption
that the main H components are at similar velocities to the metals, although
an exact quantitative match is not realized.

Model 3 incorporates a fifth hydrogen component (Fig 7c.), and
all parameters adjust to the new additional absorber.  Unlike Model 2,
the absorbers shift away from $\Delta\, v = 0$ \kms.  The total
hydrogen column is split between four main components 
the velocity
dispersions are much lower in three of the components.  
There is an improvement
in $\chi^2$, but not as significant as the addition of the fourth component.
Model 3 does not represent the whole system as well as Model 2, and
the $\chi^2$ improvement does not warrant the addition of more 
more than four components.  
Model 2 is less complex than Model 3, gives a similar
fit to the data, and agrees better with the metal lines.

In Model 4, we exclude the constraint on \nht given by the Lick spectrum,
and constrain the absorption model by the HIRES spectrum only.
The results are shown in Fig 7d. and are very similar to Fig. 7b, 
which we expect but Model 2 and 4 are identical with the exception of the
relaxed \nht constraint.  So the HIRES data alone give the same results for
D/H as those using both spectra, although the uncertainties are
larger for Model 4.  
We find that the \nht constraint is useful, but not crucial, since the 
high order Lyman lines approach unsaturation and give strong constraints
on \nht independently.  

Model 5 demonstrates the systematic offset introduced by not allowing
free parameters in the unabsorbed continuum level.
The best fits from the four previous models suggest that the continuum
level across the \Lya feature was overestimated.  Therefore Model 5
requires a higher \nht to increase the damping wings of \Lya, and 
shifts the confidence regions to lower values of D/H.  We also point
out that of all six models, the uncertainties in D/H are the smallest in
Model 5.  This is certainly not because of the \cmin of Model 5, which is the
highest of any model.  The smaller uncertainty on D/H is due to the
smaller the number of free parameters and not because Model 5 gives
a better fit to the data.

Model 6 includes an additional \hi absorber, but unlike previous models,
introduces this absorption at the position of the deuterium feature.
This unrelated \hi absorbs flux at or near the position
of \di, and represents contamination of D.  Model 6 gives the best fit
to data out of all six models.  The two strongest \hi components 
show the best match to positions of the metal lines in the 95\%
confidence region.  In Figures 8a and 8b, we show the regions of \Lya and \Lyb,
respectively, for the best fit of parameters of Model 6.   

Our main conclusion is that all six models give consistent confidence
ranges for D/H.  
In Fig. 9, we display the 95\% confidence intervals on D/H calculated in each
of the six models.
The first four models show that D/H is not very sensitive
to the number of main components nor the inclusion of the constraint
on \nht.  These first four models are consistent with 
log (D/H) = $-4.36 \pm 0.09$ (95\% confidence).
Model 5 shows that free parameters in the continua 
do affect D/H and the overall fit to
the Lyman lines.  Model 6 shows that contamination of the D is likely, and
below we discuss the implications of this result in detail.

\subsection{Contamination}

We have shown that the introduction of contamination with a single \hi
absorber gives a dramatic improvement in the best fit to the data.
The improved \cmin of Model 6 with respect to all other models
is the first indication that contamination is likely.  The effects
of the contaminating hydrogen absorber can be seen in Figure 10.
Figure 10 is similar to Fig. 7, but in Fig. 10 we show the parameters
for three \di components and the contaminating \hi absorber in Model 6.
The column densities show that the contaminating \nhi adjusts to maintain
a total column of log N $=13.1$ \cm2 for the whole deuterium feature
for low values of D/H (log (D/H) $< -4.4$).  This effectively relaxes the
constraint on the main \hi components, and allows for a better fit
to all Lyman lines for low values of D/H.  With this constraint relaxed,
the \hi components give best fit velocity positions which are very similar
to the metal line velocities, and gives more weight to the existence
of contamination.  The contaminating absorber also relaxes the constraints
on the continuum level at \Lya, and gives a better match between the 
best fit continuum level and the initial estimate.

TB realized that contamination could be likely in this absorption system,
and simulated the effect of contamination using the known distribution
of \hi absorbers at this redshift (c.f. \cite{kir97}).  TB found that
the most likely value of log (D/H) was lowered by 0.08 dex when contamination
was included, which is consistent with the results of Model 6.
TB argued that the shape of the deuterium feature has a profile which is
intrinsically too wide to be explained by deuterium alone.  We expand on this
by showing the best fit $b$-value of the contaminating absorber,
$b$(\ion{H}{1})$_{cont}$, versus D/H,
which is shown in the third panel in Fig. 10.  
$b$(\ion{H}{1})$_{cont}$ varies between 30 \kms < $b$ < 45 \kms, which is expected
of random \Lya forest lines with log N $\approx 13$ \cm2 (\cite{kir97}).

If contamination is indeed present in this D/H absorption system, then
we must attempt to determine how much contamination is expected and how
well does the expected amount of contamination fit the data.
In Figure 11, we show the \cmin function of Model 6 and a probability
that a contaminating \hi line with its minimum column density would
lie within 20 \kms of the deuterium absorption feature.  We calculated
the probability function using a power law distribution of \hi column densities 
with index $-1.5$ (\cite{kir97}, \cite{lu97}).
The convolution of the two functions gives the relative likelihood of
D/H including the expected amount of contamination (bottom panel, Fig. 10).
The horizontal dashed lines shows the likelihood level containing
95\% confidence, and the vertical dotted-dashed lines are given from the
$\chi^2$ function.  The regions are almost identical, which exemplifies that
the probability assigned for the column density of the contaminating absorber
has a very slight effect on the overall likelihood of D/H.  In other
words, the contamination is more constrained by the profile of the deuterium
feature than the probability of contamination falling near deuterium.
We may have incorrectly estimated the likelihood of contamination if the
distribution of \hi absorbers correlated with Lyman limit systems
is significantly different from the random distribution.

Therefore, with a high likelihood of contamination in this absorption
system, we adopt the confidence interval in D/H found for Model 6,
\begin{equation}
\rm{log (D/H)} = -4.40 ^{+0.06}_{-0.08},
\end{equation}
where the errors represent 67\% confidence.  

\subsection{Metal Lines}
 
An analysis of the metal absorption lines provides information on
the metallicity, ionization state, temperature and turbulent motions 
of the absorption system.

Figure 12 shows the absorption profiles of the seven detected 
metal lines in the HIRES spectrum.  Table 5 lists the 
best fit model parameters for the three components.
We included the third weaker metal line component at 
$\Delta \, v = -6$ \kms to account for the corresponding \hi component in
Model 6.  In all seven lines, we can constrain parameters
in the two stronger components (2 \& 3), but can only provide upper limits
on the column densities in the first component.  
In Table 5, we also place upper limits on other metal lines 
where no absorption was detected.

We simulate the ionization state in each component to calculate 
the metallicities.  We use a photoionizing background spectrum of 
Haardt \& Madau (1996), and the radiative transfer code, CLOUDY
(Ferland 1993).
The results from the simulations are shown in Table 6.  In
each component, we find the total hydrogen density, $n_H$,
and ionization parameter, U, which produce the observed ionic column
densities of the C and Si.
We find [C/H] = $-2.8, 3.0$ and [Si/H] = $-2.4,-2.7$ for components
2 and 3, respectively, where [X/H] is the logarithmic abundance relative
to solar.  With the derived ionization parameters, we can 
also place upper limits on N and Fe.
We find [C/Si] $\simeq -0.3$ in both components
which is characteristic of low metallicity stars in the halo of our Galaxy.
This suggests that the C and Si were created in ``normal" supernovae. If
some additional astrophysical process destroys D, it must do so without
producing more C and Si,
or other elements which we would have seen,
and without changing the C/Si ratio.  The constraints on the metallicity
and age with the characteristic size of the absorption system have
been discussed in detail by TB and Jedamzik \& Fuller (1997).

In Table 6, we also show the calculated temperatures and turbulent
velocities for components 2 and 3.  The turbulent velocity dispersion
and temperature are calculated from the observed velocity dispersions
of H, C, and Si.  We also list the equilibrium temperatures for the
ionization states calculated with CLOUDY. 
The velocity dispersions of the H and D lines 
are dominated by thermal motions, so the use of 
Voigt profiles in the fitting procedure is a justified approximation,
and we do not need to consider more complex
mesoturbulent models (\cite{lev96}).

\section{PRIMORDIAL D/H}

The measurement presented here agrees with D/H measured 
at $z=3.5722$ towards
Q1937--1009, where BT found log (D/H) = $-4.49 \pm 0.08$.  
Both systems are young and metal-poor, and D/H in each system is likely to
reflect the primordial abundance ratio.  Any process which could alter the
D abundance in either system from the primordial abundance 
would have to act identically in both systems. 
The uncertainties in both measurements are statistical, and we can
directly combine the $\chi^2$ functions from each measurement
(Fig. 13), which gives
\begin{equation}
\rm{Log (D/H)}_p = -4.47 ^{+0.030}_{-0.035}
\end{equation}
or
\begin{equation}
\rm{(D/H)}_p = 3.39 \pm 0.25 \times 10^{-5}.
\end{equation}

Many groups have calculated
the abundance yields of the light elements as a function of $\eta$
(\cite{wag67}; \cite{wal91}; \cite{smi93}; \cite{kra95}; \cite{cop95}; 
\cite{sar96}).
The abundance yield of deuterium is a single-valued function of $\eta$,
and using output from the Kawano code (\cite{kaw92}), we find
\begin{equation}
\eta = 5.1 \pm 0.3 \times 10^{-10},
\end{equation}
where the errors include
the statistical uncertainties in the D/H measurements and 
the nuclear cross-sections (\cite{sar96}).

With the present-day photon density determined from the COBE FIRAS
measurements of the Cosmic Microwave Background (\cite{fix96}),
we can directly calculate the present-day baryon density
\begin{equation}
\Omega_b \, h^2 = 0.019 \pm 0.001
\end{equation}
where $H_0 = 100 \, h$ \kms Mpc$^{-1}$.
This is a high value for $\Omega_b \, h^2$, and is consistent
with estimates of the baryon density from measurements of the \Lya forest
(\cite{rau97}; \cite{wei97}; \cite{zha97}), and galaxy clusters
(\cite{car97}).

Standard models of Galactic chemical evolution estimate the 
astration of D as a function of time and metallicity
(\cite{tyt96},\cite{edm94}),
and can account for D/H in the pre-solar nebula, 
and in the local
interstellar medium,
where D/H $=2.6 \pm 1.0 \times 10^{-5}$ (\cite{gei93})
and D/H $= 1.6 \pm 0.2 \times 10^{-5}$ (\cite{pis97}), respectively.
 
\subsection{Other Light Elements}

In Figure 14, we present the predicted abundances
of the light elements relative to hydrogen in SBBN.
The calculations assumed 3 species of 
light neutrinos and a neutron-half life of 887 s
(\cite{smi93}).  
The dashed lines represent 95\% confidence levels through
Monte Carlo simulations of reaction rate and neutron lifetime
uncertainties in the SBBN calculations (\cite{sar96}).

The boxes represent the inferred primordial abundances through
observations of D, $^4$He and $^7$Li. 
The two boxes which lie along the D abundance curve represent
D/H measurement in QSO absorption systems.  The larger box
(larger uncertainties) represents the measurement presented here
towards Q1009+2956, and the smaller represents the measurement by
BT towards Q1937--1009.  The combined result specifies 95\%
confidence levels in $\eta$, which is shown by the shaded region.

The constraint on $\eta$ given by the D/H measurement directly implies 
abundances for the other light elements.  The implied abundances
must fall in the shaded region in Fig. 14.  For example, including
SBBN uncertainties, D/H implies a primordial mass fraction of
$^4$He, 
\begin{equation}
Y_p = 0.247 \pm 0.001.
\end{equation}
This range in D/H can be compared with two recent studies of
$^4$He abundance measurements utilizing emission lines in
metal-poor extragalactic H~II regions.
Our measurements of D/H are consistent with the inferred value of
Y$_p = 0.243 \pm 0.003$ found by
Izotov et al. (1997) and shown by the larger box.  Although D/H is
not consistent with value found in another study
by Olive et al. (1997), Y$_p = 0.234 \pm 0.002$, 
shown in the smaller box.
Unlike the two measurements
constraining D/H, the determinations of Y$_p$ are not 
mutually consistent.

The bottom curve in Fig. 13 shows the predicted SBBN yields of
$^7$Li and the intersection with our confidence region from D/H.
Due to large uncertainties in the reaction rates critical to
$^7$Li, a given $\eta$ implies a large range of values for
primordial $^7$Li.  For example, our D/H measurements imply,
\begin{equation}
\rm{log ({{^7Li} \over {H}})}_p = -9.5 \pm 0.2,
\end{equation}
which is higher than abundance inferred from the Spite ``plateau''
observed in warm metal-poor halo stars (\cite{spi82}, \cite{spi84},
\cite{reb88}, \cite{tho94}, \cite{bon97}).
The latest study by Bonifacio \& Molaro (1997) use
a better indicator of effective stellar temperatures,
and infer a primordial abundance represented by the box in Fig. 14.
Even though the observational errors are similar in magnitude to our
D/H measurements (represented by the height of the boxes),
the BBN uncertainties dominate the uncertainty when determining of $\eta$.
The overlap of the box and shaded region shows the region of 
consistency between our values of D/H, and the abundance of $^7$Li
given by Bonifacio \& Molaro (1997).  Although, primordial $^7$Li
could be higher by as much as 0.6 dex, remain consistent with
the D/H measurements, and accommodate non-standard models of
lithium depletion (\cite{vau95}, \cite{pis92}).

We conclude that our D/H measurements 1) are likely the first measurements
of a primordial abundance ratio of any element, 2) give the best determination 
of $\eta$ from SBBN and 3) are consistent with current observations of the other
light elements.

\section{IMPLICATIONS FOR IBBN}

The measurements of low D/H in two separate high-redshift QSO absorption
systems has immediate consequences for inhomogeneous models of Big
Bang nucleosynthesis (IBBN).  Jedamzik et al. (1993) first pointed out that
observational constraints on the primordial value of D/H can directly
test for and constrain the spatial variations of $\eta$ during the
epoch of BBN.  The amount of deuterium produced by BBN is very
sensitive to both the average entropy density {\it and} the spatial
variations in entropy.  A generic feature of all IBBN models is that
deuterium production is enhanced over standard homogeneous
models (\cite{jed94a}).  IBBN models which include either sub-horizon (\cite{jed94b})
or super-horizon (\cite{jed95}) entropy fluctuations typically predict deuterium
production which is 10 times greater than SBBN.
Because deuterium production is so sensitive to $\eta$, high
entropy regions are the dominant contributors to the cosmological
deuterium abundance, even though these same regions contribute only
a small fraction of the total mass.

We present two typical examples of IBBN in the context of the light
element abundances of D, $^4$He, and $^7$Li.  In the first
example, we assume that the $\Omega_b$ is comparable to the value
we found in the previous section, $\Omega_b \, h^2 = 0.02$.
IBBN models can readily produce the inferred primordial abundance
ratios of $^4$He and $^7$Li, but predict that the mass weighted cosmic
average D/H $> 10^{-4}$.  Our two low measurements clearly rule this
out if they represent the cosmic mean D/H.  
In the second example, we choose a much higher $\Omega_b$, so that 
the cosmic mean average in IBBN models can easily accommodate 
our measurements: $\Omega_b \, h^2 = 0.15$.  All models of IBBN
overproduce $^4$He and $^7$Li, with typical predictions of
Y$_p$ = 0.26 and $^7 \rm{Li}/\rm{H} = 10^{-9}$, which lie well
beyond the upper limits inferred from extragalactic H~II regions
(\cite{oli97}; \cite{izo97}) and population II halo stars (\cite{bon97}),
respectively.

The simplest interpretation is that the low D/H measurements
are consistent with the other light elements with predictions of
homogeneous models of BBN.  The assumption of SBBN is justified
with a low primordial value of D/H.  Following the conclusions
of Jedamzik et al. (1994), we believe that our measurements
of D/H not only provide observational constraints on 
$\eta$ and $\Omega_b$, but place limits on small scale entropy
fluctuations during the epoch of BBN.

\acknowledgments
We are extremely grateful to W.M. Keck foundation which made this
work possible, to Steve Vogt and his team for building HIRES, and 
to Tom Bida, Randy Campbell and Wayne Wack for 
assistance at the telescope.  We are very grateful to Joseph Miller
and the staff of Lick Observatory for the construction, maintenance,
and assistance with the Kast spectrograph, which was made possible by
a generous gift from William and Marina Kast.
We thank 
Christian Cardall, George Fuller, Karsten Jedamzik,
David Kirkman, Martin Lemoine, and Jason X. Prochaska
for many useful conversations.

\clearpage

%****************** STANDARD TABLE *******************************
% for centering Table 1

\begin{table*}
\tablenum{1}
\begin{center}
\begin{tabular}{ccccc}
\tableline
\tableline
Date & Exposure (s) & XD Order\tablenotemark{a} &
$\lambda_{min}$ (\AA) & $\lambda_{max}$ (\AA) \cr
\tableline
28 Dec 95 & 9000 & 1 & 3579.0 & 5529.0 \cr
28 Dec 95 & 7200 & 1 & 3579.0 & 5529.0 \cr
28 Dec 95 & 4800 & 2 & 3165.0 & 4330.0 \cr
09 Dec 96 & 7900 & 2 & 3135.0 & 4386.0 \cr
10 Dec 96 & 7800 & 2 & 3135.0 & 4386.0 \cr
10 Dec 96 & 8000 & 2 & 3135.0 & 4386.0 \cr
\tableline
\end{tabular}
\end{center}
\caption{HIRES Observations of Q1009+2956}
\tablenotetext{a}{Order of Cross-Disperser. First order observations
used 1x2 on-chip binning, and second order used 1x4.  All observations
used a 1.14'' slit, giving a spectral resolution of 8\kms FWHM}
\end{table*}

%\clearpage

\begin{table*}
\tablenum{2}
\begin{center}
\begin{tabular}{cccccc}
\tableline
\tableline
Region & $\lambda_{min}$ & $\lambda_{max}$ &
Pixels & Order\tablenotemark{a} & SNR\tablenotemark{b} \cr
\tableline
\Lya & 4254.50 & 4264.00 & 332 & 5 & 60 \cr
\Lyb & 3591.00 & 3596.60 & 231 & 3 & 25 \cr
Ly-6 & 3259.18 & 3262.20 & 155  & 3 & 13 \cr
Ly-12 -- Ly-14  & 3208.00 & 3214.60 & 306 & 2 & 10 \cr
Ly-Limit & 3199.40 & 3202.00 & 119 & 1  & 6 \cr
\tableline
\end{tabular}
\end{center}
\caption{Spectral Regions used in D/H Measurement }
\tablenotetext{a}{Order of Legendre polynomial used for the continuum}
\tablenotetext{b}{Approximate Signal-to-Noise Ratio at continuum level
per 2 \kms pixel}
\end{table*}

%\clearpage
\begin{table*}
\tablenum{3}
\begin{center}
\begin{tabular}{ccc}
\tableline
\tableline
log N (\cm2) & $b$(\kms) & $z$ \cr
\tableline
12.65   & 9.6  & 1.63915 \cr
12.90   & 18.6 & 1.64226 \cr
13.53   & 75.3 & 1.64351 \cr
12.79   & 21.9 & 1.95511 \cr
13.31   & 19.7 & 1.95473 \cr
12.57   & 39.0 & 1.95722 \cr
13.14   & 28.8 & 2.50076 \cr
13.57   & 39.0 & 2.50456 \cr
\tableline
\end{tabular}
\end{center}
\caption{Extra H~I Lines in D/H Models }
\end{table*}

%\clearpage
\begin{table*}
\tablenum{4}
\begin{center}
\begin{tabular}{ccccccc}
\tableline
\tableline
Model & Com\tablenotemark{a}  & D/H ($-2\sigma$)\tablenotemark{b} &
D/H($\chi^2_{min}$) & D/H ($+2\sigma$)\tablenotemark{b} & $\chi^2_{min}$ &
n - $\nu$\tablenotemark{c} \cr
\tableline
 1  & 3 & $-4.44$ & $-4.37$ & $-4.28$ & 942.6 & 51 \cr
 2  & 4 & $-4.45$ & $-4.37$ & $-4.29$ & 906.1 & 55 \cr
 3  & 5 & $-4.43$ & $-4.35$ & $-4.26$ & 892.2 & 59 \cr
 4\tablenotemark{d} & 4 & $-4.44$ & $-4.34$ & $-4.27$ & 903.5 & 55 \cr
 5\tablenotemark{e} & 4 & $-4.52$ & $-4.45$ & $-4.39$ & 1076.1 & 41 \cr
 6\tablenotemark{f} & 4 & $-4.56$ & $-4.40$ & $-4.28$ & 828.2 & 58 \cr
\tableline
\end{tabular}
\end{center}
\caption{D/H Absorption Models}
\tablenotetext{a}{Number of main components in fit}
\tablenotetext{b}{95\% confidence levels from $\chi^2$ test}
\tablenotetext{c}{Number of free parameters}
\tablenotetext{d}{\nht~ constraint is not included }
\tablenotetext{e}{continua are not allowed to vary}
\tablenotetext{f}{Contamination included at D-\Lya}
\end{table*}

%\clearpage
\begin{table*}
\tablenum{5}
\begin{center}
\begin{tabular}{ccccccc}
\tableline
\tableline
Ion  & \multispan{2}{\hfil Comp. 1 \hfil}
& \multispan{2}{\hfil Comp. 2 \hfil}
& \multispan{2}{\hfil Comp. 3 \hfil} \cr
  & \multispan{2}{\hfil $(z=2.50351)$ \hfil}
& \multispan{2}{\hfil $(z=2.50357)$ \hfil}
& \multispan{2}{\hfil $(z=2.50370)$ \hfil} \cr
 &  Log N & b & Log N & b & Log N & b \cr
\tableline
\hi & $15.79 \pm 0.02$ & $21.1 \pm 8.1$ & $16.93 \pm 0.02$ & $15.3 \pm 3.0$
    & $17.07 \pm 0.01$ & $18.6 \pm 5.7$ \cr
\di & 11.39  & $16.1 \pm 7.9$ & 12.53 & $11.6 \pm 3.0$
    & 12.67 & $17.5 \pm 5.5$ \cr
\noalign{\vskip 15pt}

\ion{C}{1} & $< 12.2 $\tablenotemark{a}  & ...
    & $<12.0 $\tablenotemark{a}  & ...
    & $<12.0 $\tablenotemark{a}  & ... \cr
\ion{C}{2} & $< 12.4 $\tablenotemark{a}  & ...
    & $12.30 \pm 0.13$ & $4.6 \pm 2.3$
    & $12.18\pm 0.13$ & $3.9 \pm 2.4$ \cr
\ion{C}{3} & $< 12.3 $\tablenotemark{a}  & ...
    & $13.31 \pm 0.07$ & $4.7 \pm 0.8$
    & $13.44 \pm 0.04$ & $9.4 \pm 2.4$ \cr
\ion{C}{4} & $< 12.1 $\tablenotemark{a}  & ...
    & $12.81 \pm 0.02$ & $5.4 \pm 0.3$
    & $12.56 \pm 0.03$ & $5.6 \pm 0.6$ \cr
\noalign{\vskip 15pt}

\ion{N}{1} & $< 12.6 $\tablenotemark{a}  & ...
    & $<12.3 $\tablenotemark{a}  & ...
    & $<12.3 $\tablenotemark{a}  & ... \cr
\ion{N}{2} & $< 12.7 $\tablenotemark{a}  & ...
    & $<12.4 $\tablenotemark{a}  & ...
    & $<12.4 $\tablenotemark{a}  & ... \cr
\ion{N}{5} & $< 12.5 $\tablenotemark{a}  & ...
    & $<12.4 $\tablenotemark{a}  & ...
    & $<12.4 $\tablenotemark{a}  & ... \cr
\noalign{\vskip 15pt}

\ion{O}{1} & $< 12.9 $\tablenotemark{a}  & ...
 & $< 13.0 $\tablenotemark{a}  & ...
 & $< 12.8 $\tablenotemark{a}  & ... \cr
\noalign{\vskip 15pt}

\ion{Si}{2} & $< 11.8 $\tablenotemark{a}  & ...
 & $< 11.6$\tablenotemark{a}  & ...
 & $< 11.5$\tablenotemark{a}  & ... \cr
\ion{Si}{3} & $< 11.7$\tablenotemark{a}  & ...
    & $12.79 \pm 0.05$ & $4.6 \pm 0.3$
    & $12.59 \pm 0.02$ & $4.5 \pm 0.8$ \cr
\ion{Si}{4} & $< 12.0$\tablenotemark{a}  & ...
    & $12.47 \pm 0.02$ & $4.0 \pm 0.4$
    & $12.05 \pm 0.03$ & $3.9 \pm 0.7$ \cr
\noalign{\vskip 15pt}

\ion{Fe}{2} & $< 12.7$\tablenotemark{a}  & ...
    & $<12.4$\tablenotemark{a}  & ...
    & $<12.1$\tablenotemark{a}  & ... \cr
\ion{Fe}{3} & $< 12.8$\tablenotemark{a}  & ...
    & $<12.5$\tablenotemark{a}  & ...
    & $<12.9$\tablenotemark{a}  & ... \cr
\tableline
\end{tabular}
\end{center}
\caption{Column Densities of Metals}
\tablenotetext{a}{2$\sigma$ upper limits}
\end{table*}

%\clearpage
\begin{table*}
\tablenum{6}
\begin{center}
\begin{tabular}{cccc}
\noalign{\vskip 5pt}
\tableline
 & Component 1\tablenotemark{a} & Component 2 & Component 3 \cr
\tableline
[C/H]    & $< -2.8$ &   $-2.8$ &   $-3.0$ \cr
[N/H]    & $< -0.4$ & $< -1.7$ & $< -1.8$ \cr
[Si/H]   & $< -2.1$ &   $-2.4$ &   $-2.7$ \cr
[Fe/H]   & $< 0.8$  & $< -0.9$ & $< -0.7$ \cr

Log U     & ... & $-2.48$ & $-2.58$ \cr
Log \hi/H & ... & $-2.97$ & $-2.84$ \cr

Log $n_{\rm{H}}$\tablenotemark{b} (cm$^{-3}$) & ... & $-2.30$ & $-2.20$ \cr
L(kpc)  & $> 0.1$ & 5.0 & 3.0 \cr

T$_{b}$\tablenotemark{c} ($10^4$K) & $\approx 2.2$  & 1.2 $\pm$ 0.3 & 2.1 $\pm$
0.5 \cr
T$_{e}$\tablenotemark{d} ($10^4$K) & ... & 2.2 & 2.1 \cr
b$_{tur}$ (\kms) & $\approx 8.5$ & 4.8 $\pm$ 0.8 & 1.9 $\pm$ 0.9 \cr
\tableline
\end{tabular}
\end{center}

\caption{Metallicity and Ionization State}
\tablenotetext{a}{Metallcities are calculated with ionization parameter
of Component 2}
\tablenotetext{a}{Corresponding to Log $J_0 = -21.45$}
\tablenotetext{b}{Determined from component line widths}
\tablenotetext{c}{Photoionization equilibrium temperature}

\end{table*}

\clearpage

\begin{figure}
\figurenum{1}
\centerline{
\psfig{file=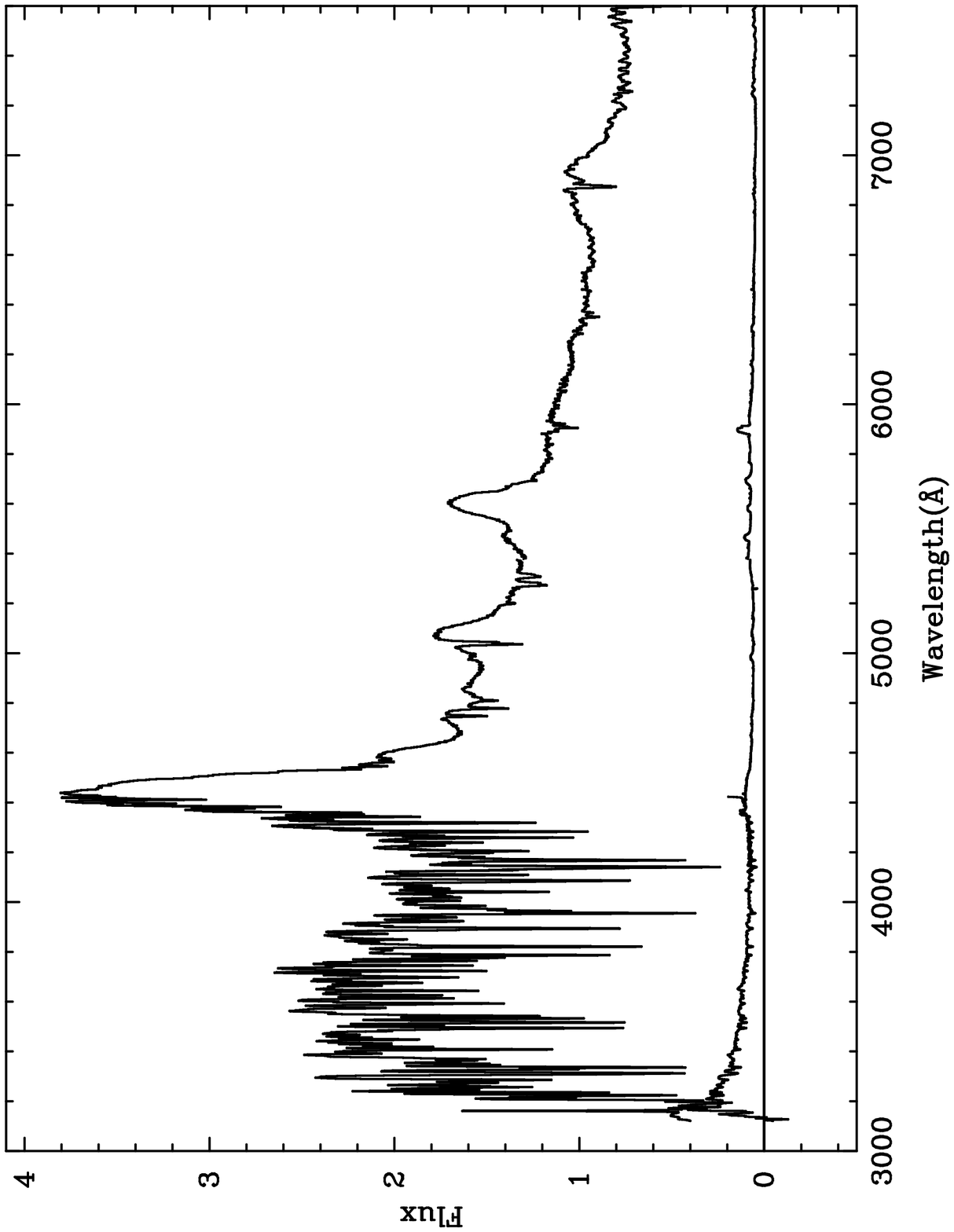,height=7.25 in,width=\columnwidth}}
\caption{
Wide slit, flux-calibrated Lick spectra of Q1009+2956
($z_{em} = 2.63$, V=16.0).  The figure contains three separate spectra
of Q1009+2956 with different setups, and the flux calibration agrees between
all three spectra.
The 10$\sigma$ is shown as the solid line near 0.1.
The Lyman limit of $z=2.504$ is the
break in the flux near 3200 \AA.
}
\end{figure}

\clearpage
\begin{figure}
\figurenum{2}
\centerline{
\psfig{file=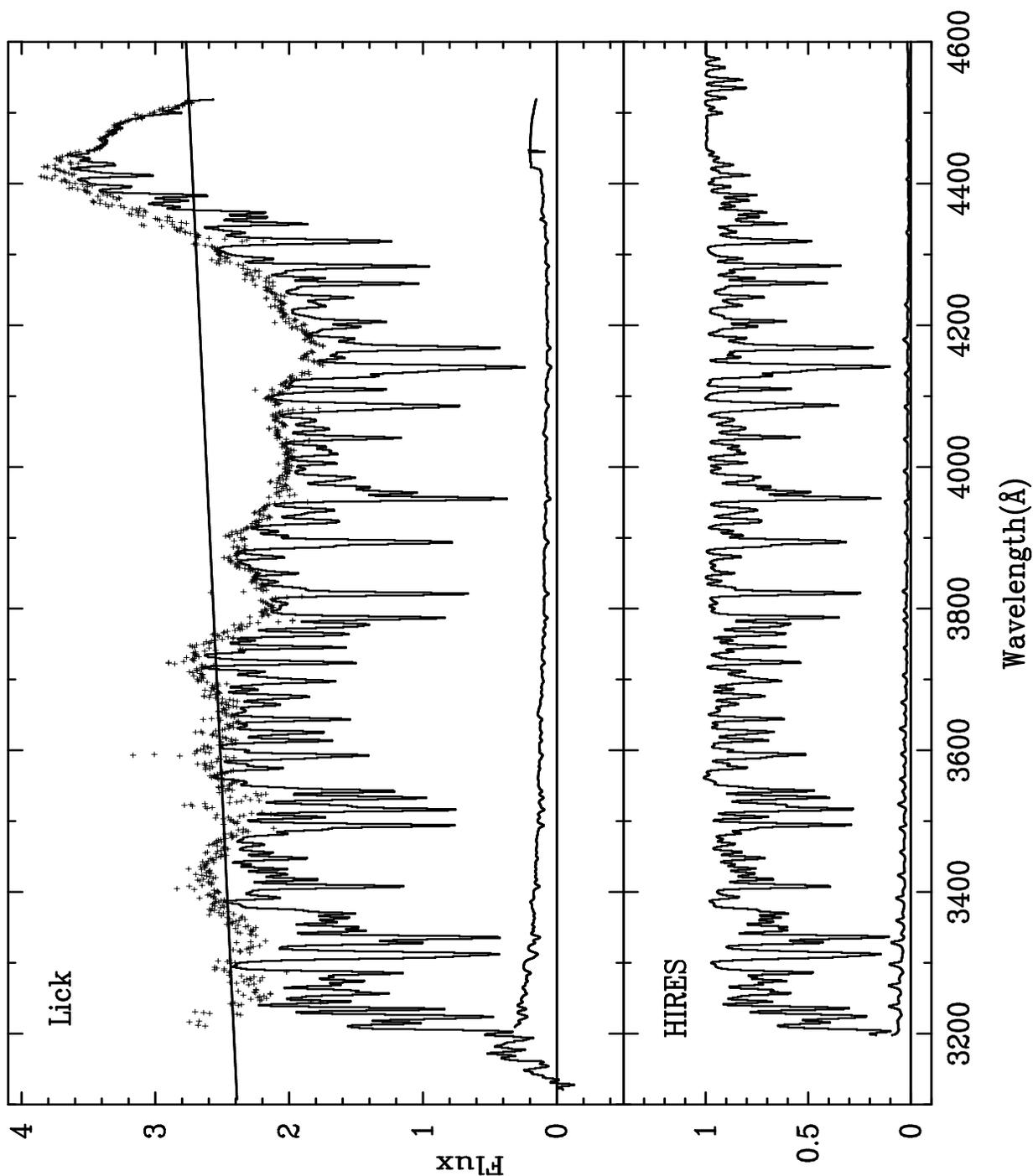,height=7.25 in,width=\columnwidth}}
\caption{
The \Lya forest of Q1009+2956 in the Lick spectrum (top panel)
and the smoothed HIRES spectrum (bottom panel).  The crosses in the top
panel show the result of dividing the HIRES spectrum into the Lick
spectrum, and represent the QSO continuum determined in the HIRES spectrum.
The straight line is the fit to the crosses below 3750 \AA (1030\AA rest),
which we use to extrapolate the continuum below 3200 \AA.
The 10$\sigma$ is shown as the solid line near 0.2.
}
\end{figure}
\clearpage

\begin{figure}
\figurenum{3}
\centerline{
\psfig{file=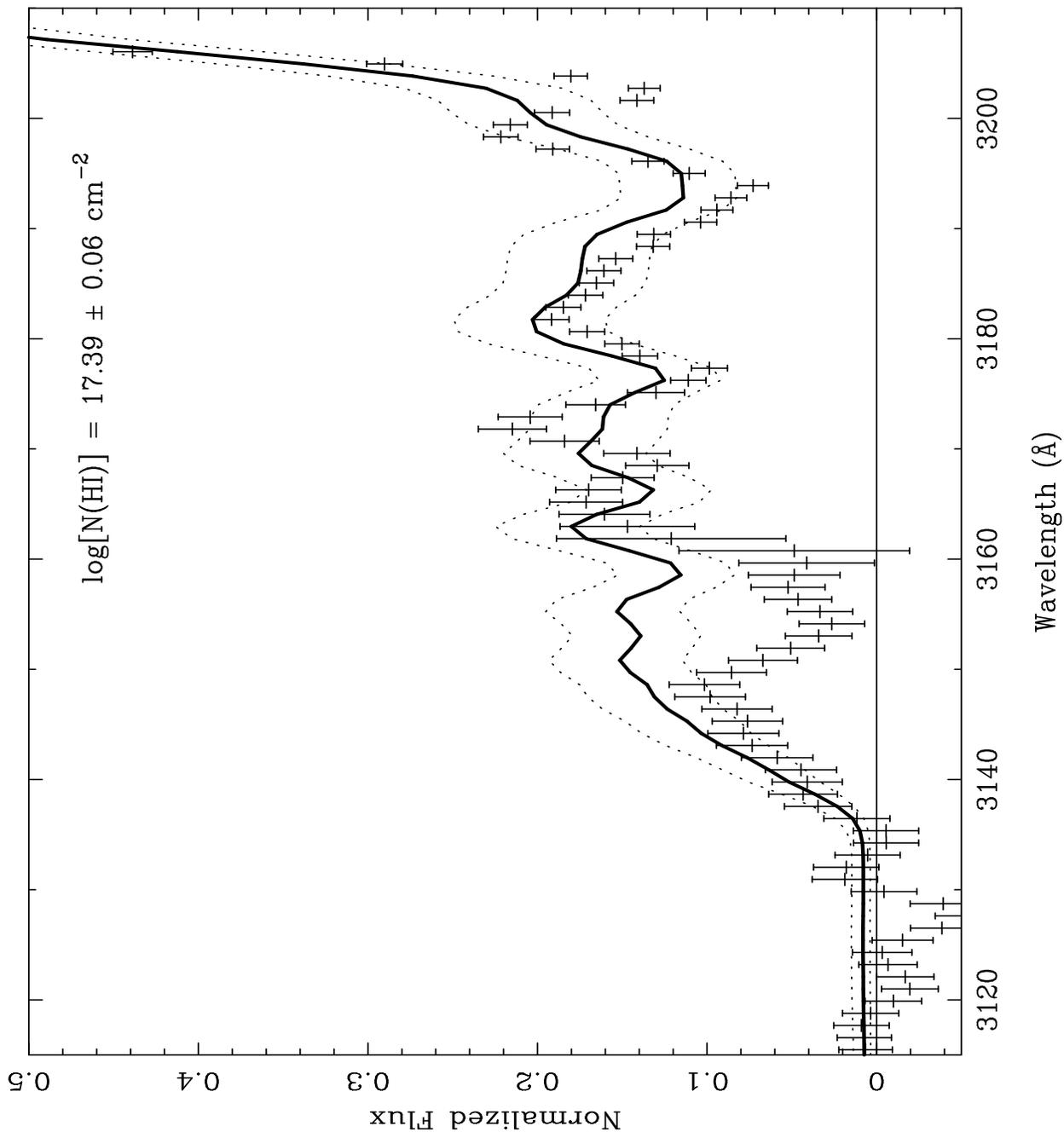,height=7.25 in,width=\columnwidth}}
\caption{
The Lick spectrum below the Lyman limit.  The flux in shown
in each pixel with 1$\sigma$ error bars.  The solid line shows the
model absorption profile with log[N(H~I)$_{total}$] = 17.39 cm$^{-2}$ 
and additional
absorption lines of H~I with $z > 2.0$.  There is another Lyman limit
system at $z=2.430$, with its break at $\lambda = 3140$ \AA.
The dotted line corresponds to the 1$\sigma$ error in the measurement
of N(H~I)$_{total}$.
}
\end{figure}
\clearpage

\begin{figure}
\figurenum{4}
\centerline{
\psfig{file=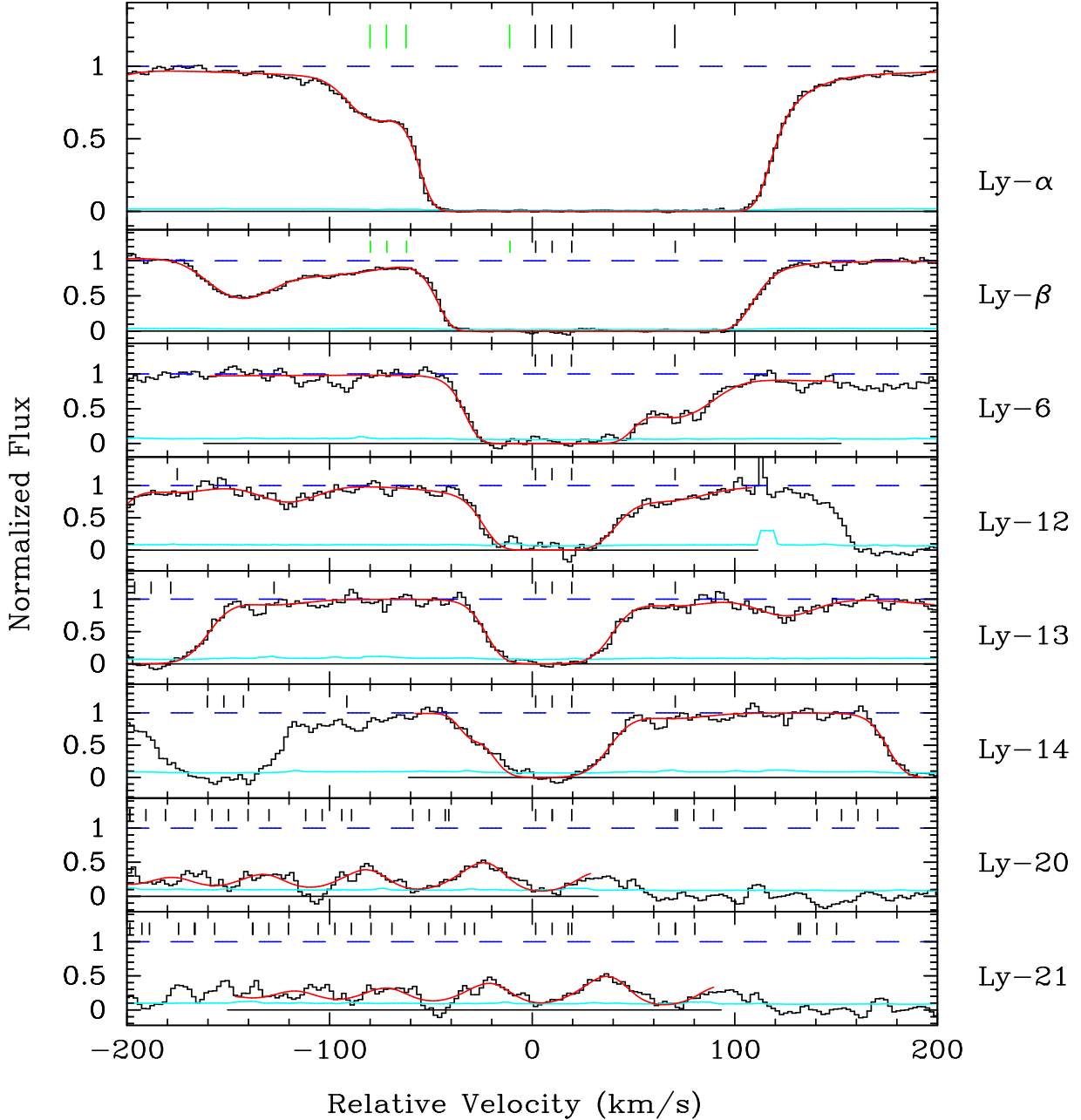,height=7.25 in,width=\columnwidth}}
\caption{
HIRES spectrum of Lyman series lines of the DHAS
stacked in velocity space.
Zero velocity corresponds to redshift $z = 2.503571$.
The histogram represents the
normalized flux, each bin corresponds to a 2 \kms pixel.
The 1$\sigma$ error is gray solid line near zero.
The smooth black line shows the best fit of Model 2.
The ticks mark the velocity positions of individual components.
This model has four main components, with three H~I lines near 0 \kms,
three D~I near $-82$ \kms, and a fourth component of H~I near
72 \kms.
The solid black line at zero flux shows the regions about each
Lyman line used in the fitting procedure.
}
\end{figure}
\clearpage

\begin{figure}
\figurenum{5}
\centerline{
\psfig{file=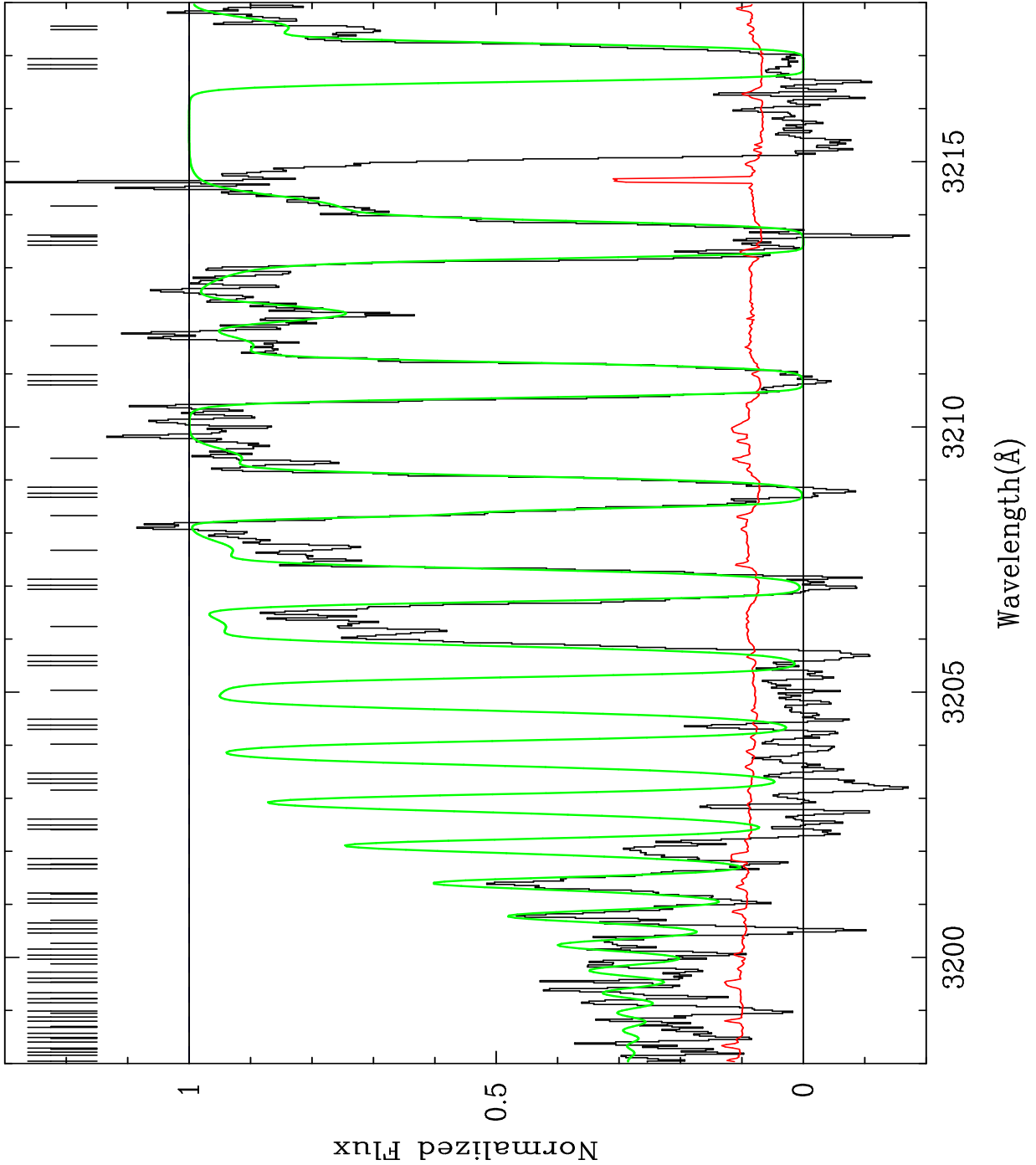,height=7.25 in,width=\columnwidth}}
\caption{
The Lyman limit region of the HIRES spectrum containing Lyman lines of Ly-11 to
Ly-24.
The histogram represents the observed flux in each 2\kms pixel normalized to the
 initial
estimate of the QSO unabsorbed continuum.  The 1$\sigma$ error level in each pix
el is
represented by the solid line near 0.1.  The smooth grey line represents the bes
t parameter fit
of Model 2.  The Lyman line centers are unsaturated above Ly-16, which gives a g
ood
constraint on the total N(H~I)$_{total}$
in this system from the HIRES spectrum alone.
}
\end{figure}
\clearpage

\begin{figure}
\figurenum{6a}
\centerline{
\psfig{file=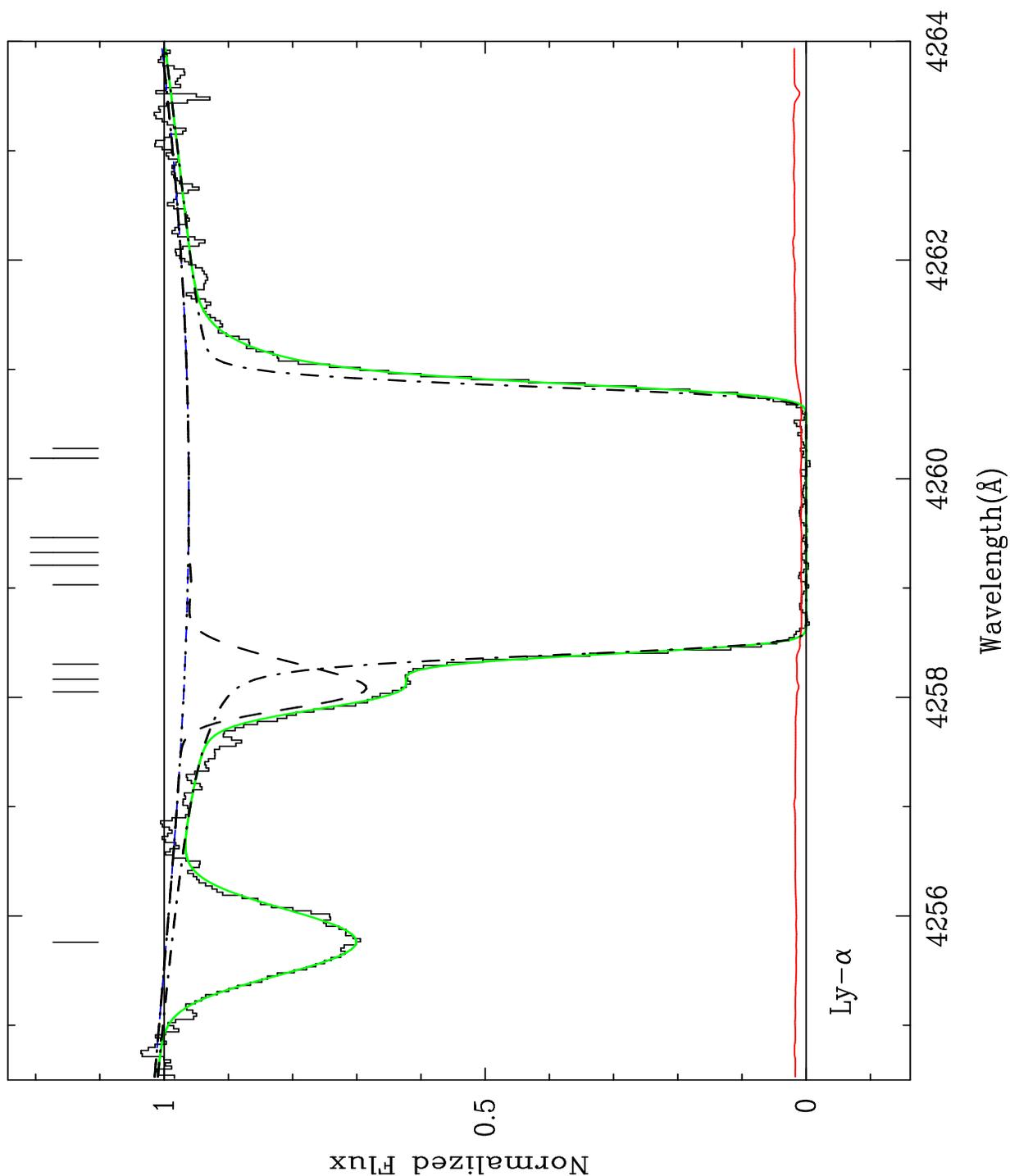,height=7.25 in,width=\columnwidth}}
\caption{
The \Lya absorption feature at $z=2.504$.  The histogram
shows the same spectrum as in Fig. 4.  The spectrum has been normalized
to the initial continuum estimate, which is the solid black line at unity.
The best fit continuum is the 5th order Legendre polynomial which drops below un
ity.
The profile fits of D and H are shown as dashed and dot-dashed lines,
respectively.  The tallest tick marks show the line centers of the
three main H components in Model 2.  The corresponding D components
are the three tick marks to the left, and the additional hydrogen
component is the lone tick mark to the right.  The grey line represents
the total absorption model fit.
}
\end{figure}
\clearpage

\begin{figure}
\figurenum{6b}
\centerline{
\psfig{file=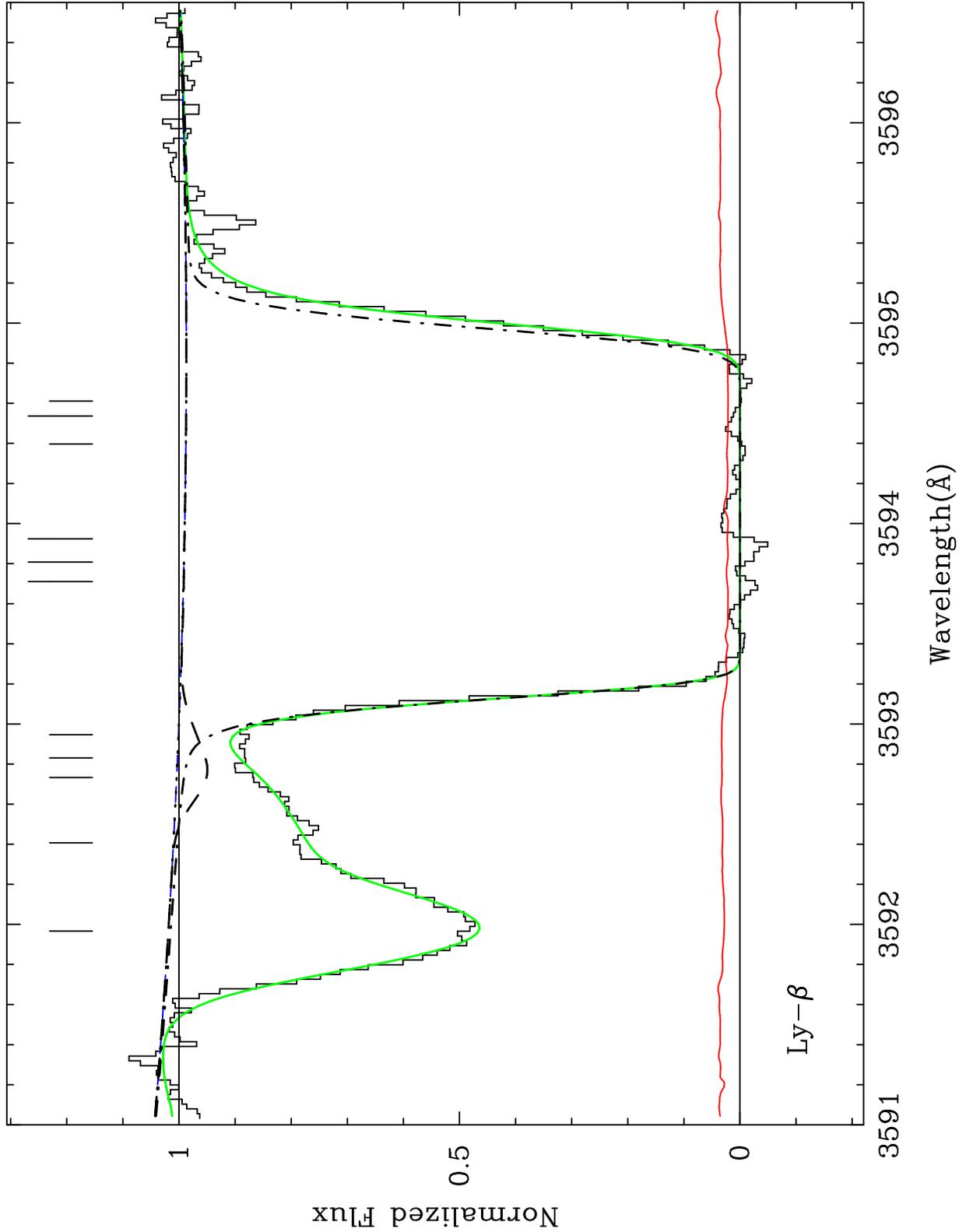,height=8.25 in,width=\columnwidth}}
\caption{
HIRES Spectrum of the \Lyb region of the DHAS.  The continuum
is a 3rd order Legendre polynomial.
Extra H~I absorption can be seen left of the position of D-\Lyb.
The gray line represents the same model as in Fig 6a.
}
\end{figure}
\clearpage

\begin{figure}
\figurenum{6c}
\centerline{
\psfig{file=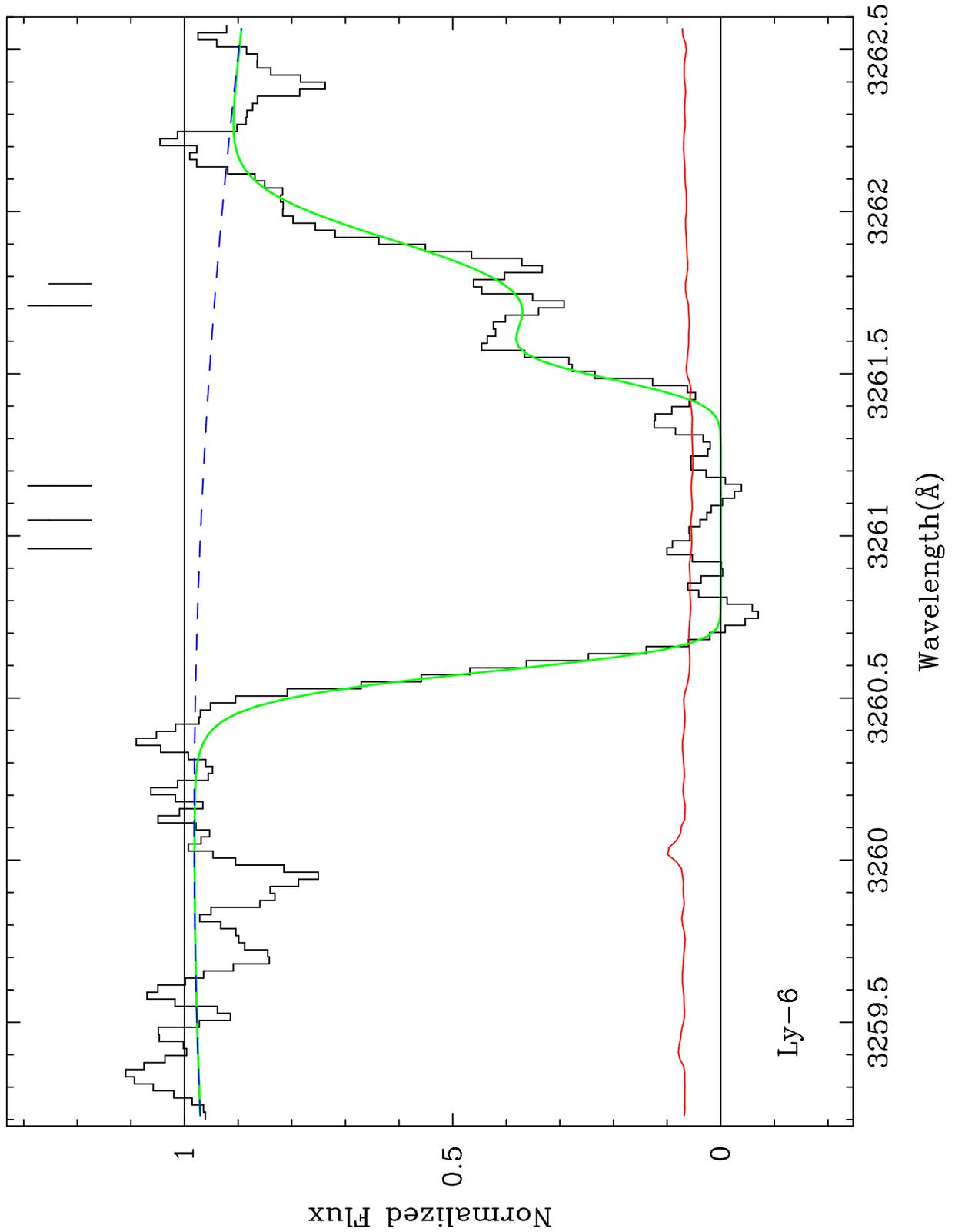,height=8.25 in,width=\columnwidth}}
\caption{
HIRES Spectrum of the Ly-6 region of the DHAS.  The continuum
is a 3rd order Legendre polynomial.
The gray line represents the same model as in Fig 6a. 
}
\end{figure}
\clearpage

\begin{figure}
\figurenum{6d}
\centerline{
\psfig{file=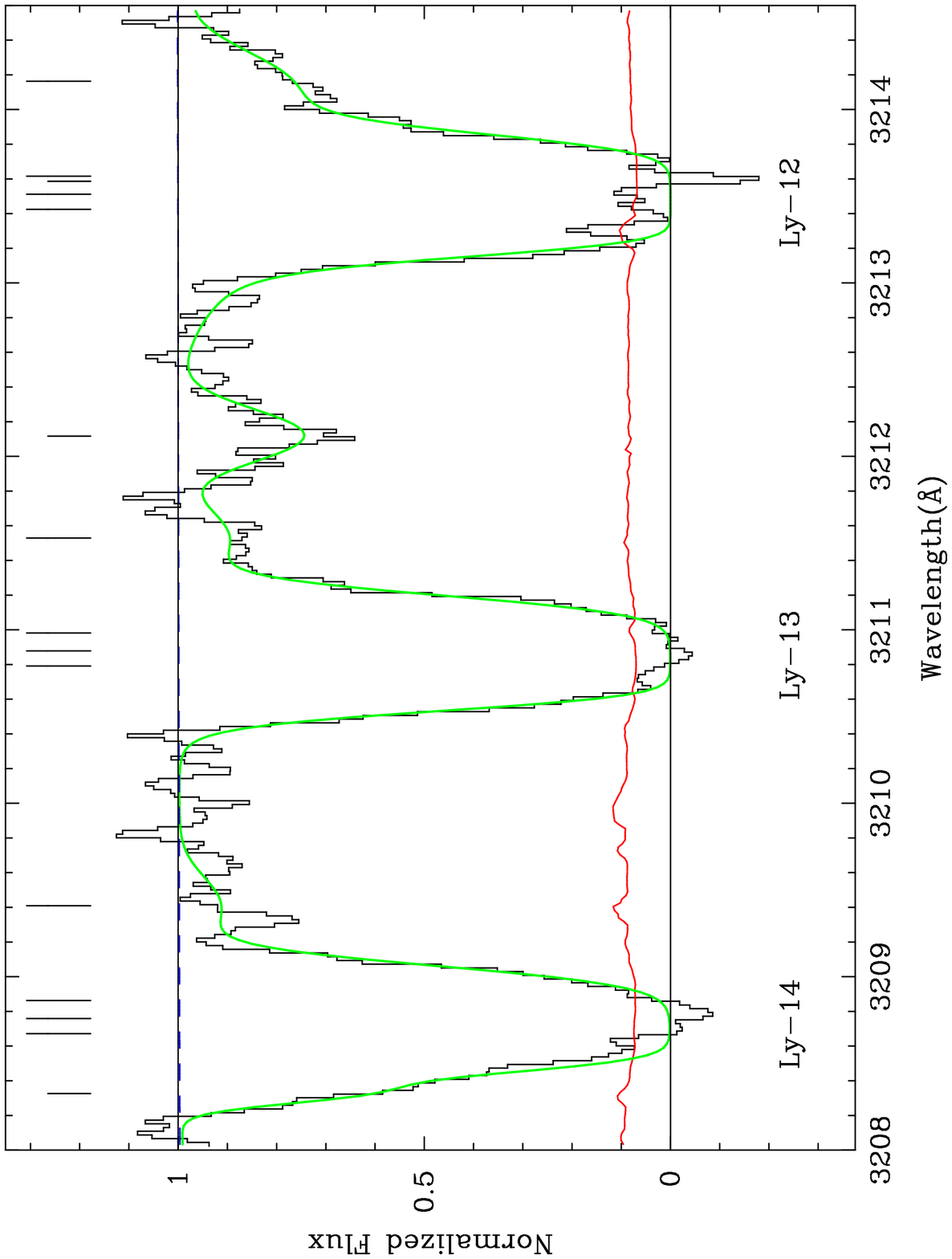,height=8.25 in,width=\columnwidth}}
\caption{
HIRES Spectrum of the region containing Ly-12 to Ly-14 of the DHAS.
The continuum
is a 2nd order Legendre polynomial.
The gray line represents the same model as in Fig 6a. 
}
\end{figure}
\clearpage

\begin{figure}
\figurenum{6e}
\centerline{
\psfig{file=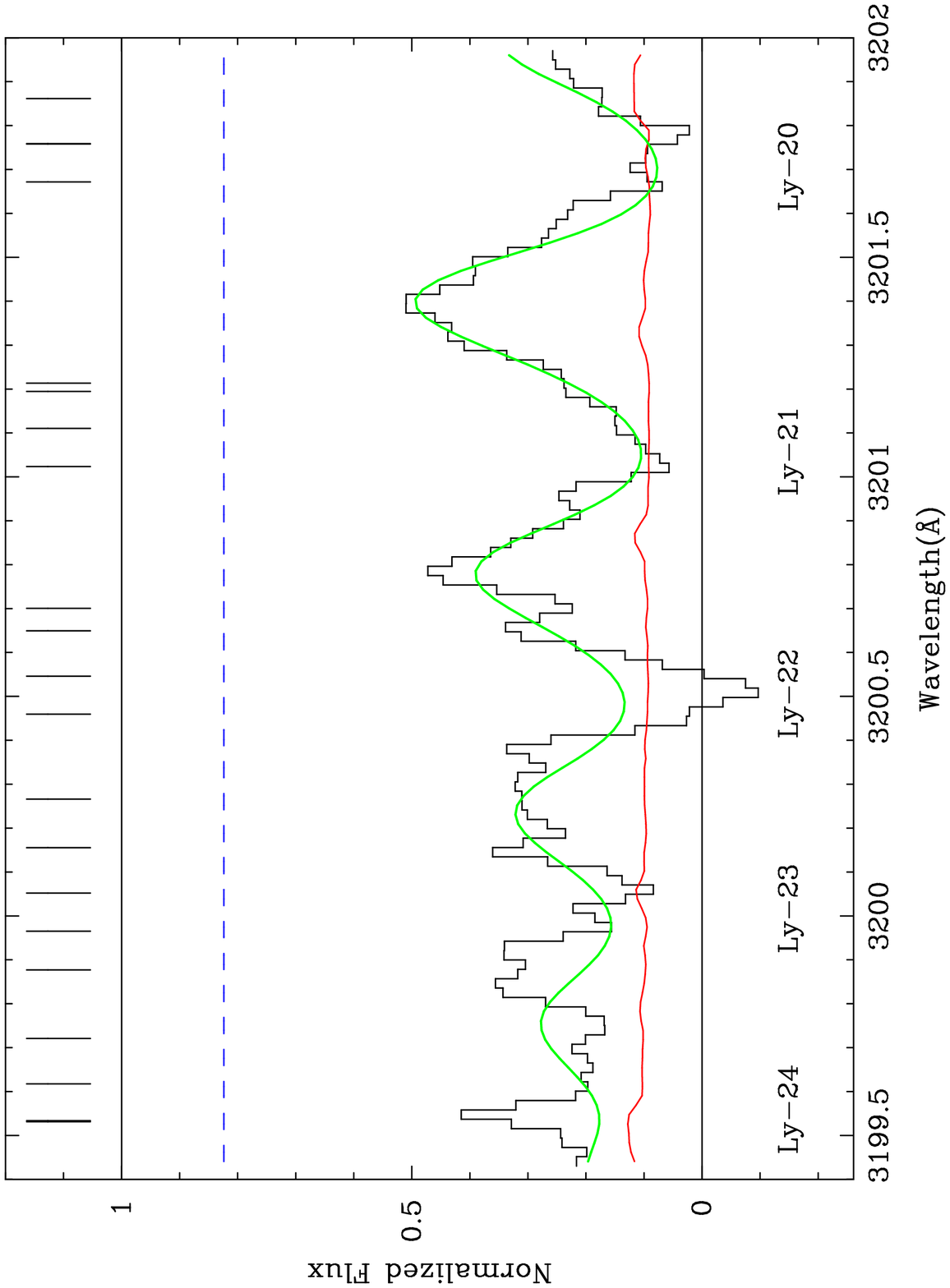,height=8.25 in,width=\columnwidth}}
\caption{
HIRES Spectrum of the region containing Ly-20 to Ly-24 of the DHAS.  The continu
um
is a 1st order Legendre polynomial.
The gray line represents the same model as in Fig 6a. 
}
\end{figure}
\clearpage

\begin{figure}
\figurenum{7a}
\centerline{
\psfig{file=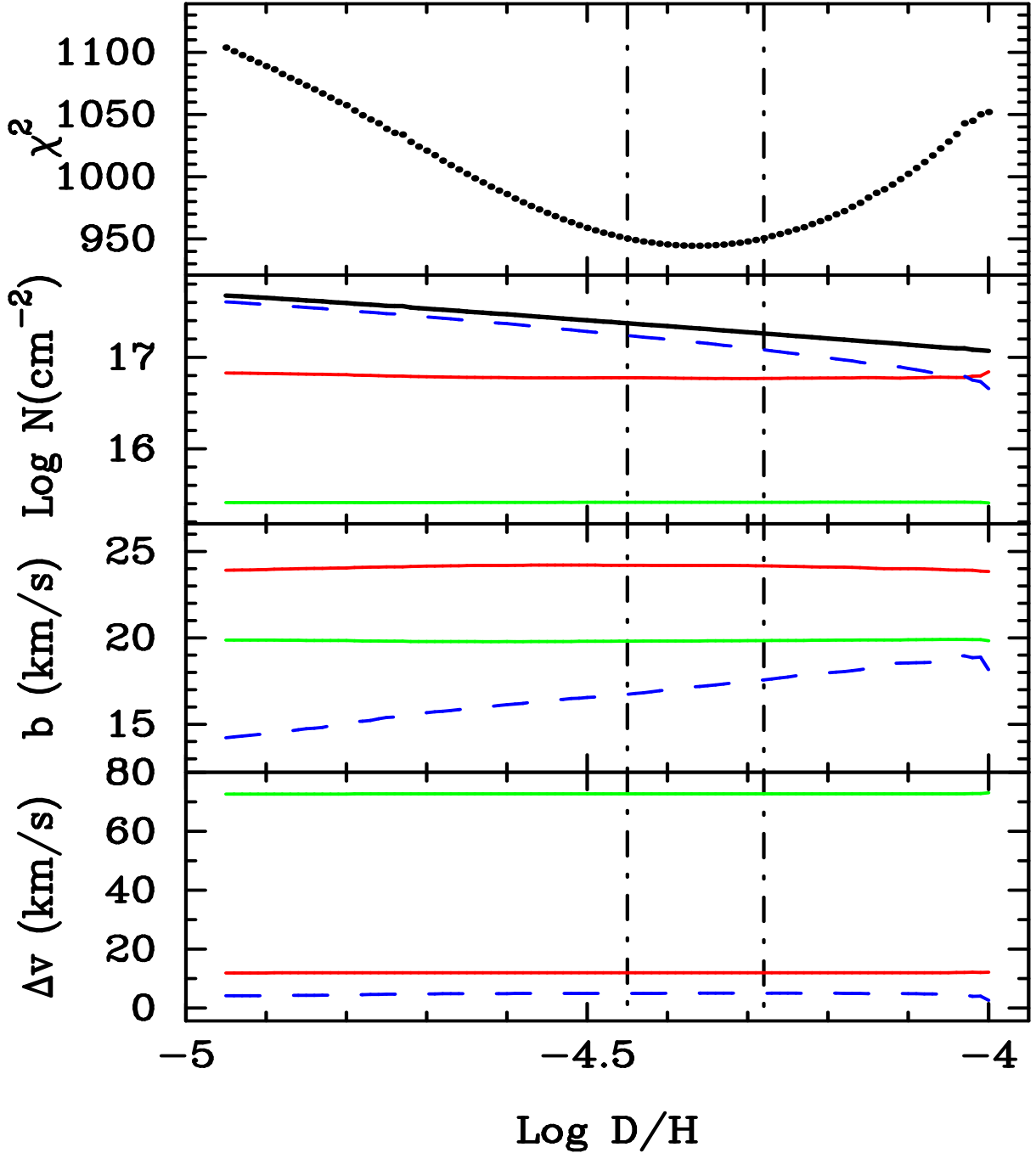,height=7.25 in,width=\columnwidth}}
\caption{
Results of the fitting procedure for Model 1.
The top panels show the \cmin functions
vs. D/H, and the vertical dot-dashed lines encompass the minima and represent
95\% confidence levels in each model.  The second panel shows the
column densities of main H~I components, with each component's value
represented with a different line style.  The thick solid line shows
N(H~I)$_{total}$ (the sum of column densities) as a function of D/H.
The third panel shows the velocity dispersions of the hydrogen components,
and the fourth shows the the velocity position relative to $z=2.503571$.
}
\end{figure}
\clearpage

\begin{figure}
\figurenum{7b}
\centerline{
\psfig{file=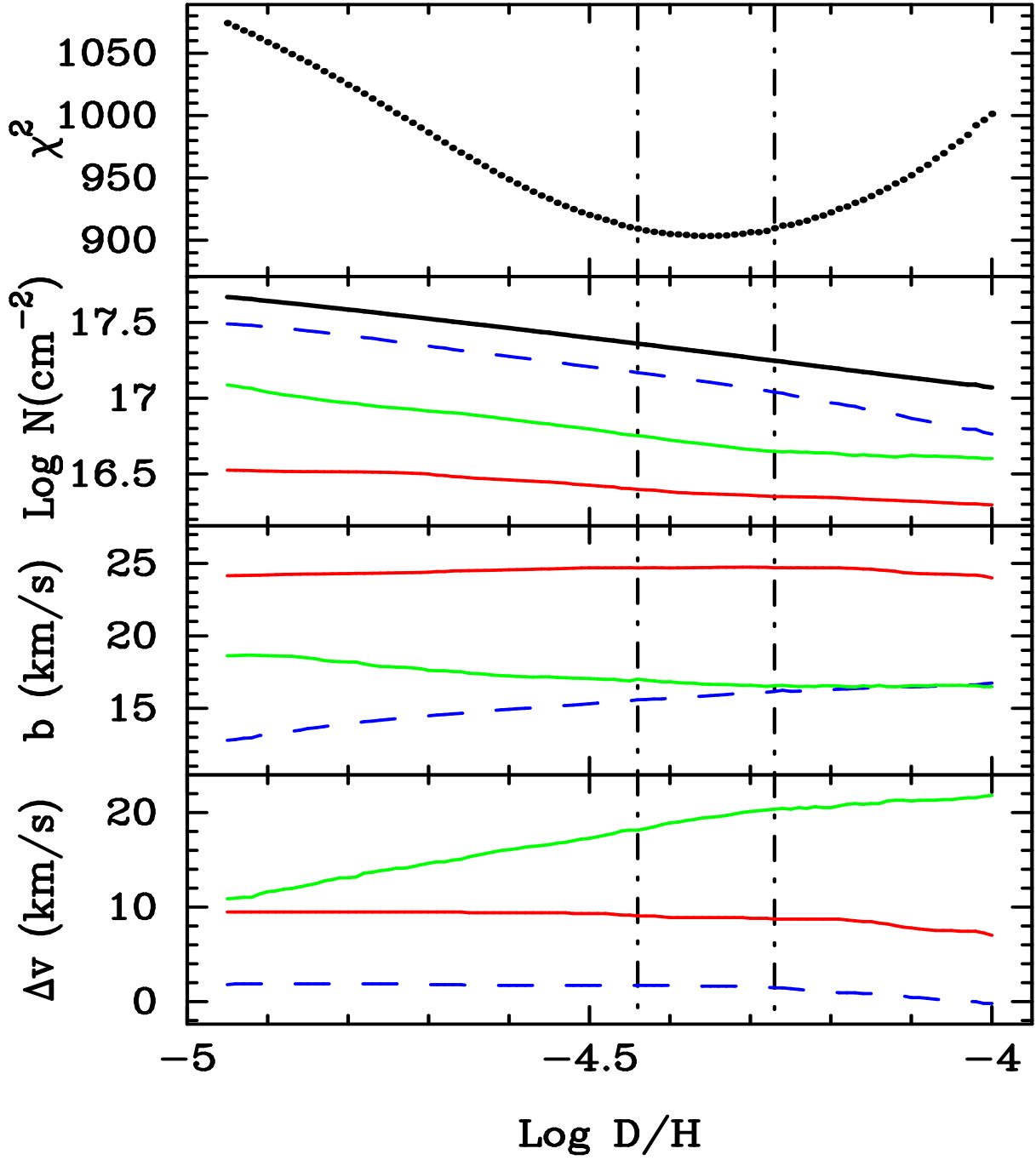,height=7.25 in,width=\columnwidth}}
\caption{
Results of Model 2 
}
\end{figure}
\clearpage

\begin{figure}
\figurenum{7c}
\centerline{
\psfig{file=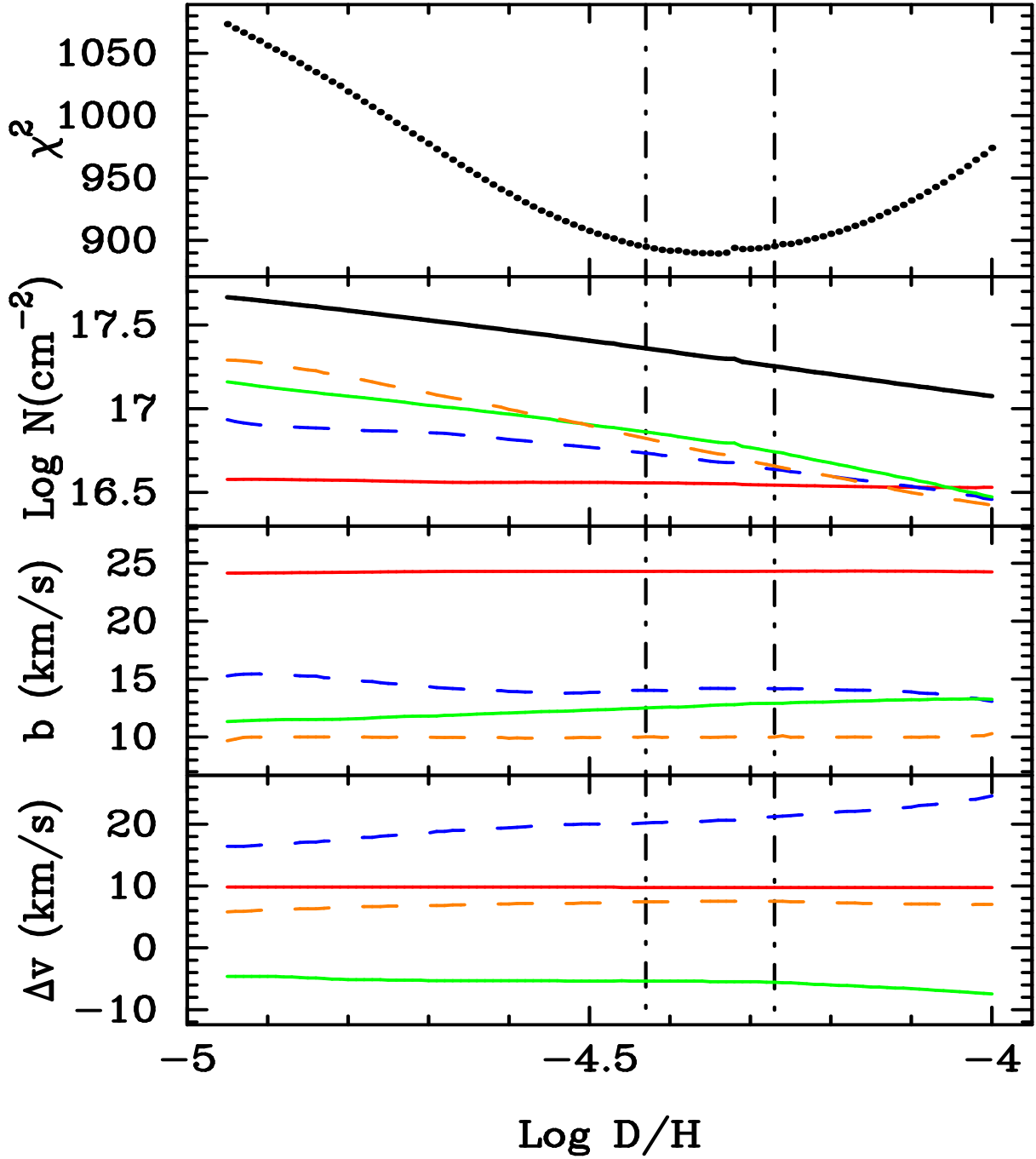,height=7.25 in,width=\columnwidth}}
\caption{
Results of Model 3 
}
\end{figure}
\clearpage

\begin{figure}
\figurenum{7d}
\centerline{
\psfig{file=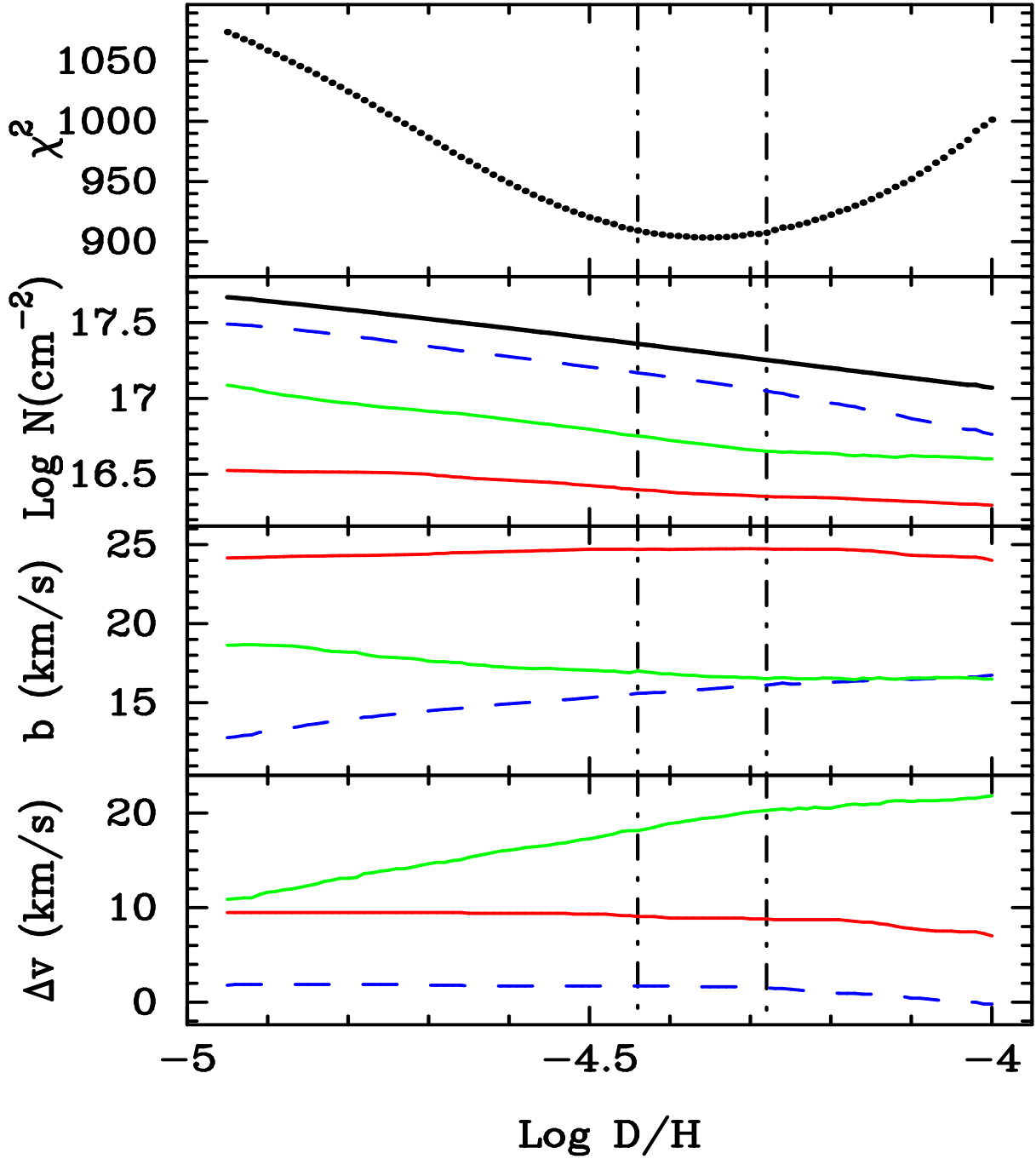,height=7.25 in,width=\columnwidth}}
\caption{
Results of Model 4 
}
\end{figure}
\clearpage

\begin{figure}
\figurenum{7e}
\centerline{
\psfig{file=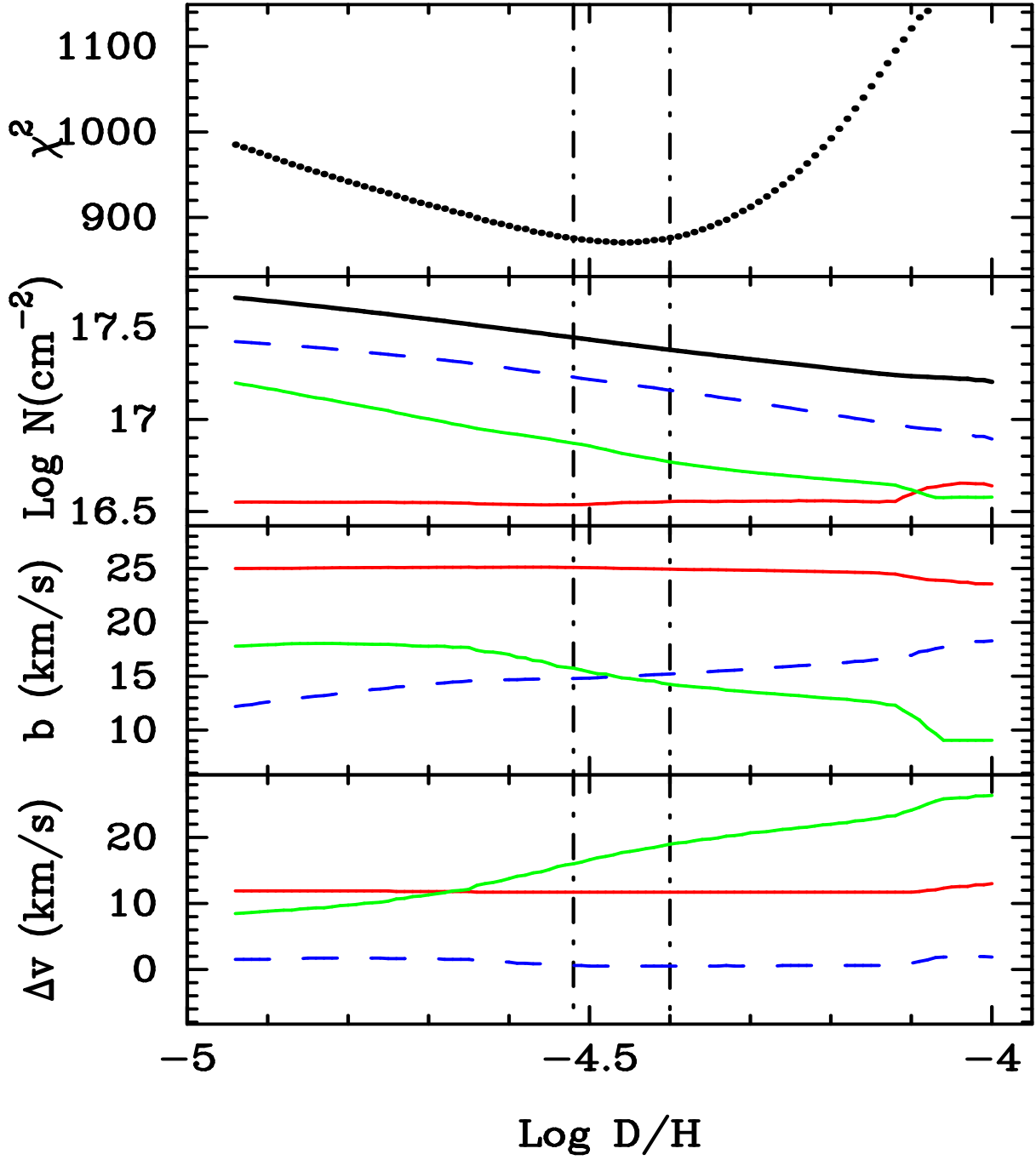,height=7.25 in,width=\columnwidth}}
\caption{
Results of Model 5 
}
\end{figure}
\clearpage

\begin{figure}
\figurenum{7f}
\centerline{
\psfig{file=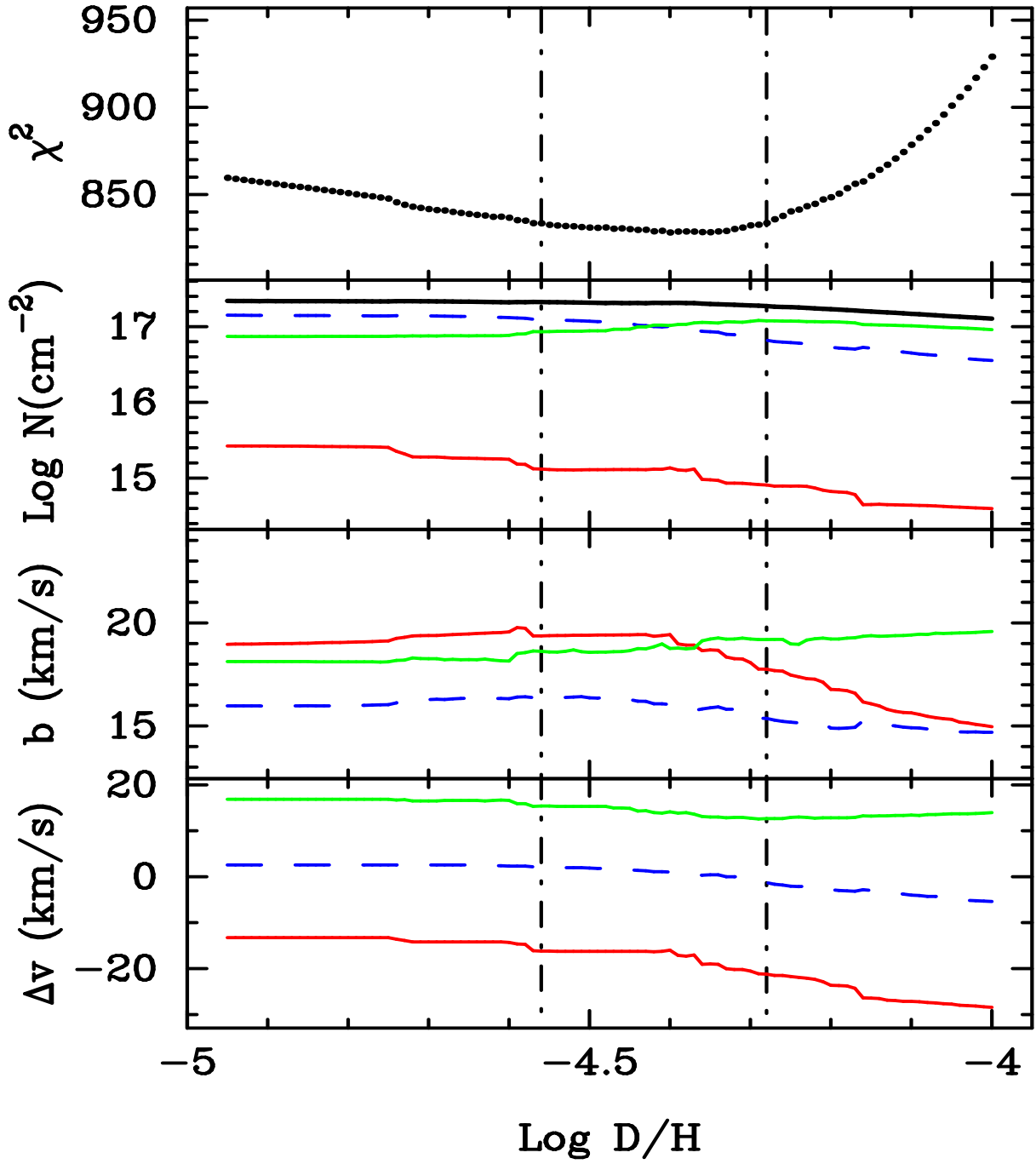,height=7.25 in,width=\columnwidth}}
\caption{
Results of Model 6 
}
\end{figure}
\clearpage

\begin{figure}
\figurenum{8a}
\centerline{
\psfig{file=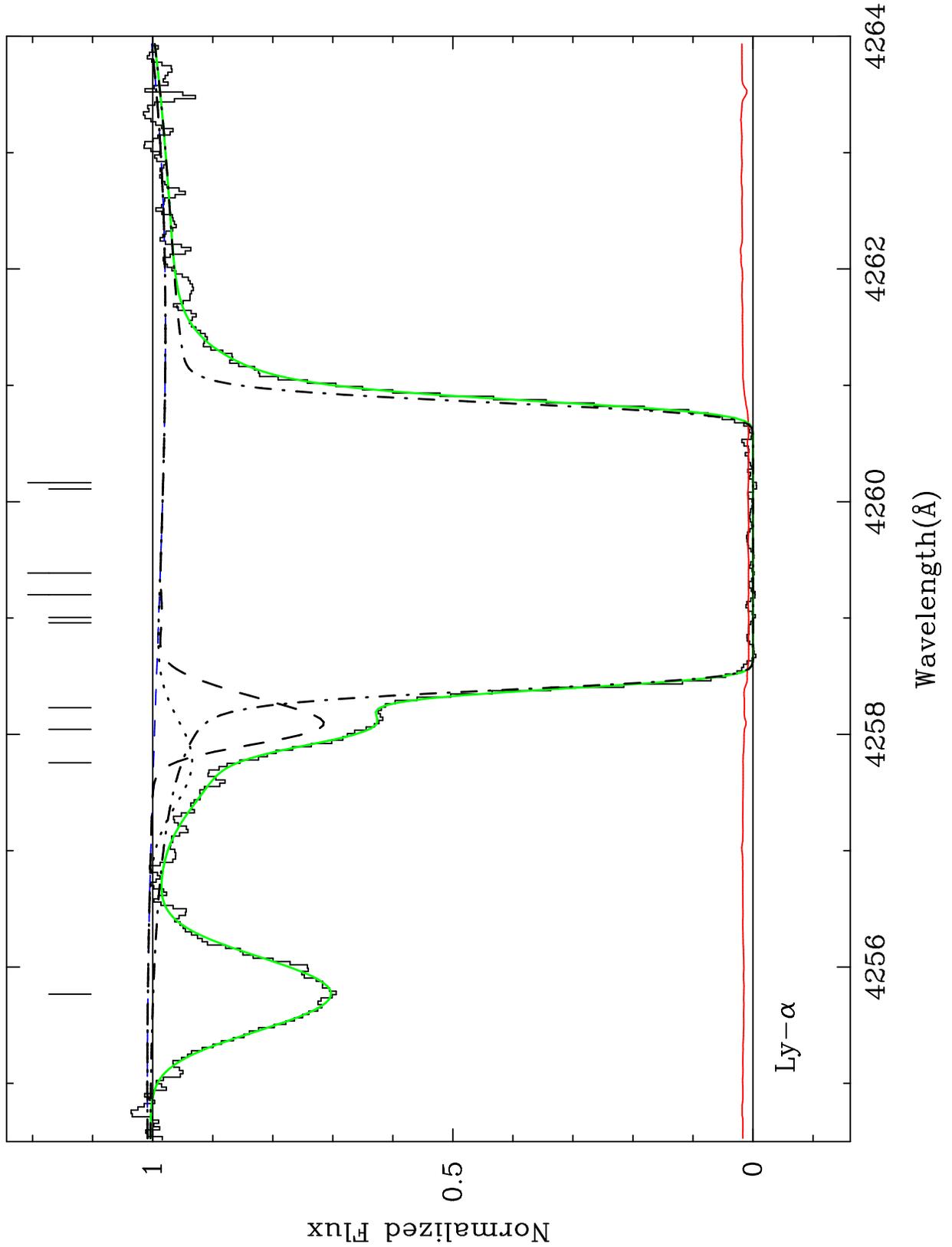,height=8.25 in,width=\columnwidth}}
\caption{
Same as Fig. 6a, but shows the best fits for Model 6 (with contamination).
The dotted line shows the profile of the H~I contamination near
the deuterium feature. 
}
\end{figure}
\clearpage

\begin{figure}
\figurenum{8b}
\centerline{
\psfig{file=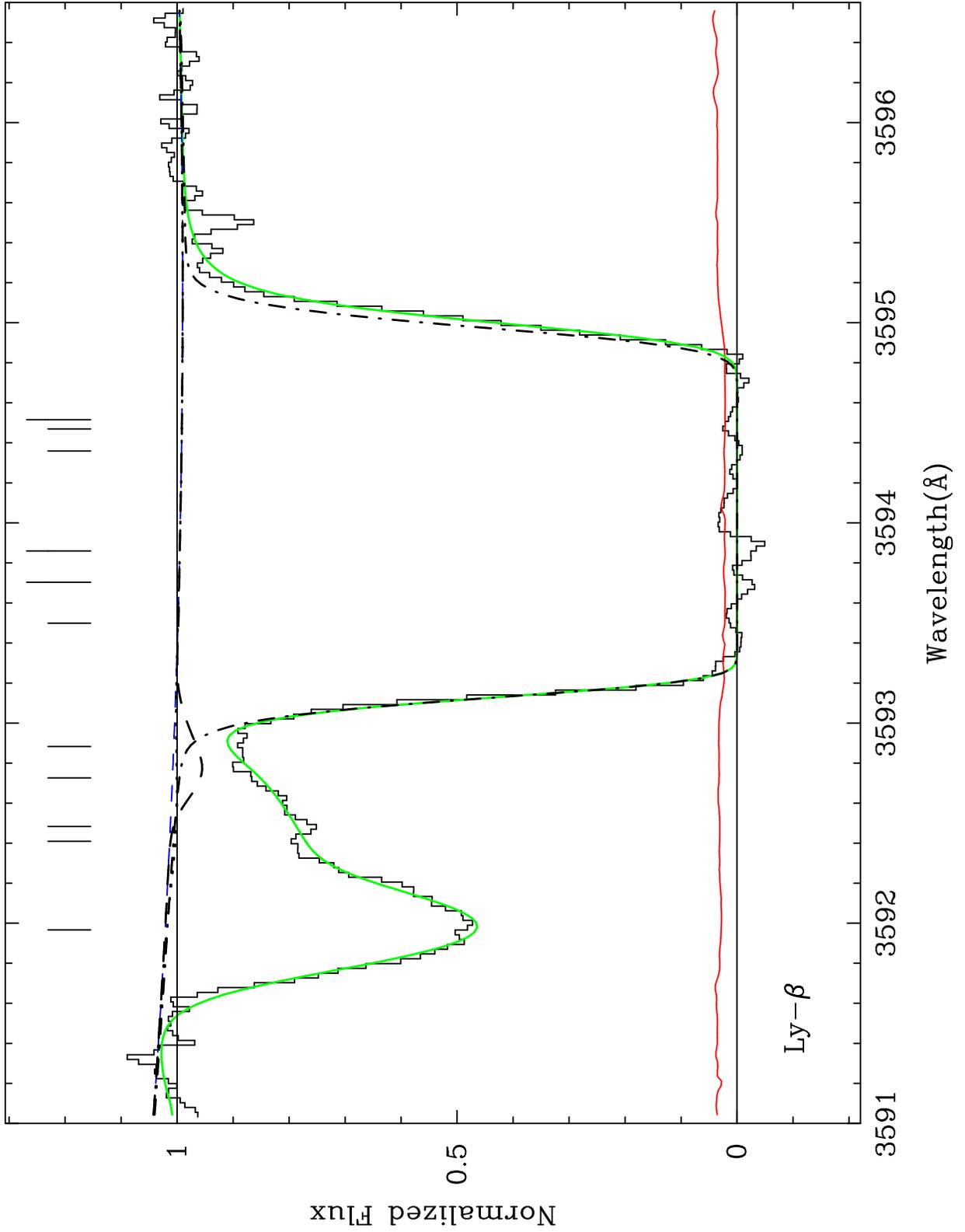,height=8.25 in,width=\columnwidth}}
\caption{
Same as Fig. 8a, but for \Lyb. 
}
\end{figure}
\clearpage

\begin{figure}
\figurenum{9}
\centerline{
\psfig{file=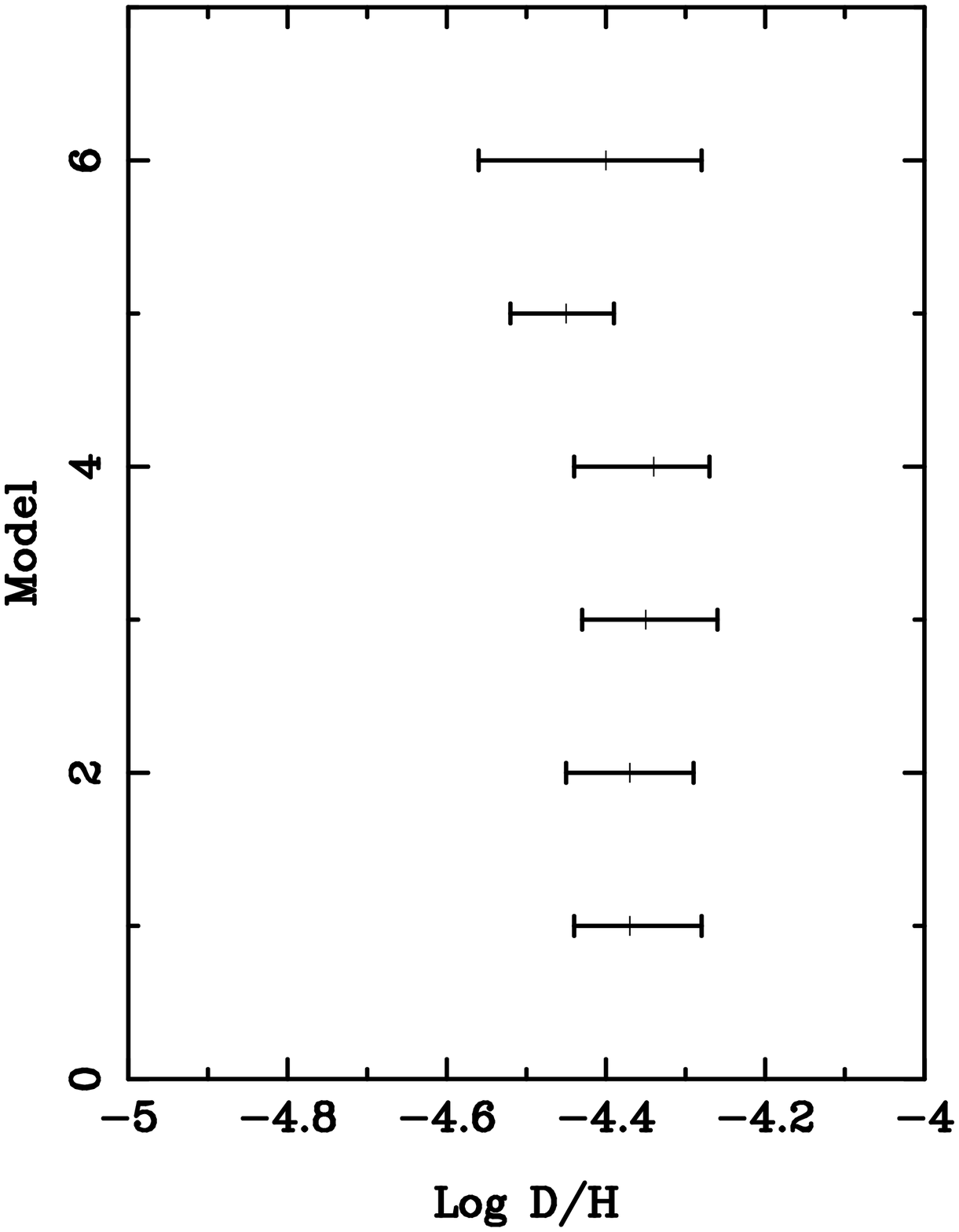,height=8.25 in,width=\columnwidth}}
\caption{
Summary of 95\% confidence intervals in the six models 
}
\end{figure}
\clearpage

\begin{figure}
\figurenum{10}
\centerline{
\psfig{file=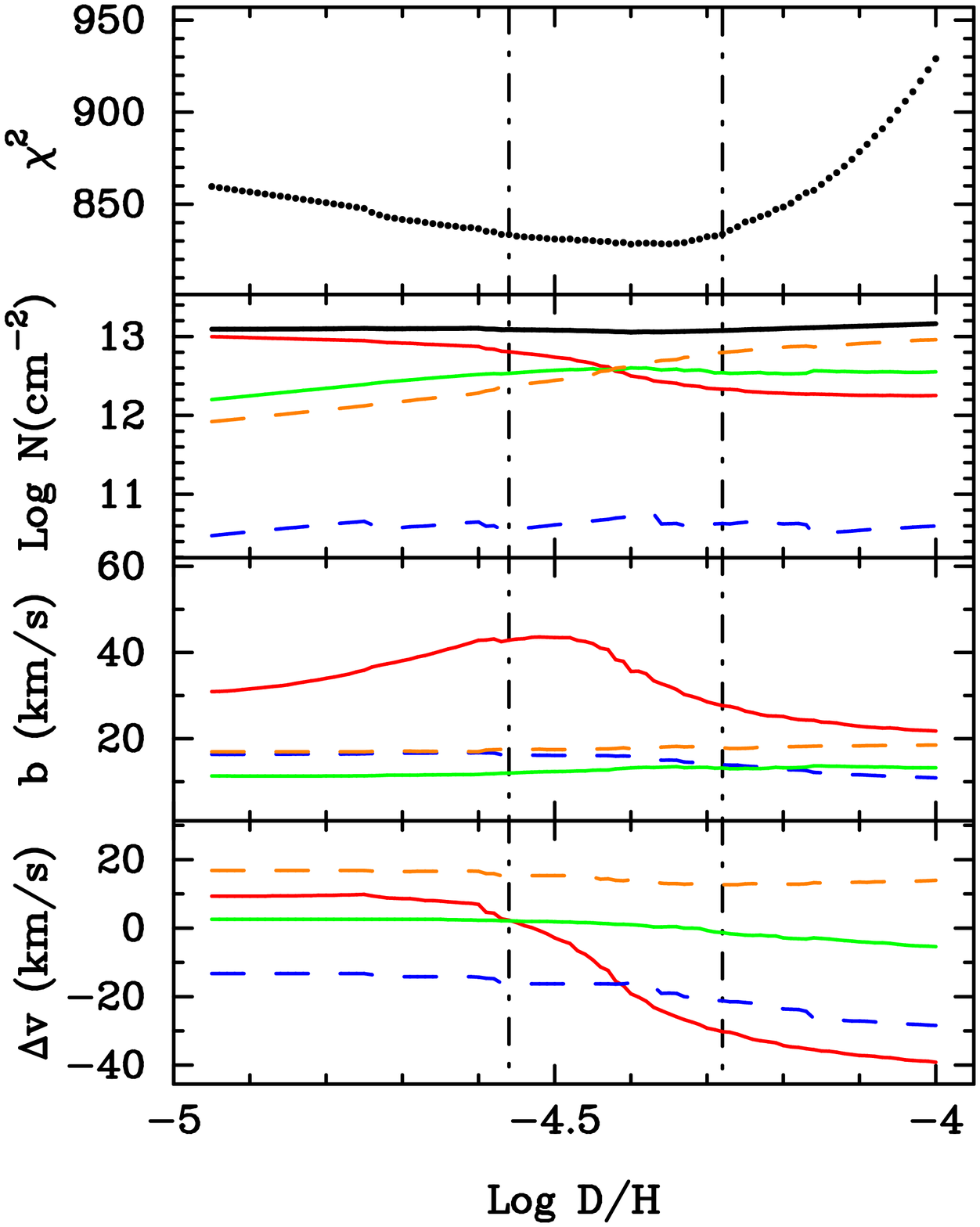,height=8.25 in,width=\columnwidth}}
\caption{
Results of the fitting procedure of Model 6.  The parameters
are displayed as in Fig. 7f, but represent the components of D~I and
the contaminating H~I absorber. 
}
\end{figure}
\clearpage

\begin{figure}
\figurenum{11}
\centerline{
\psfig{file=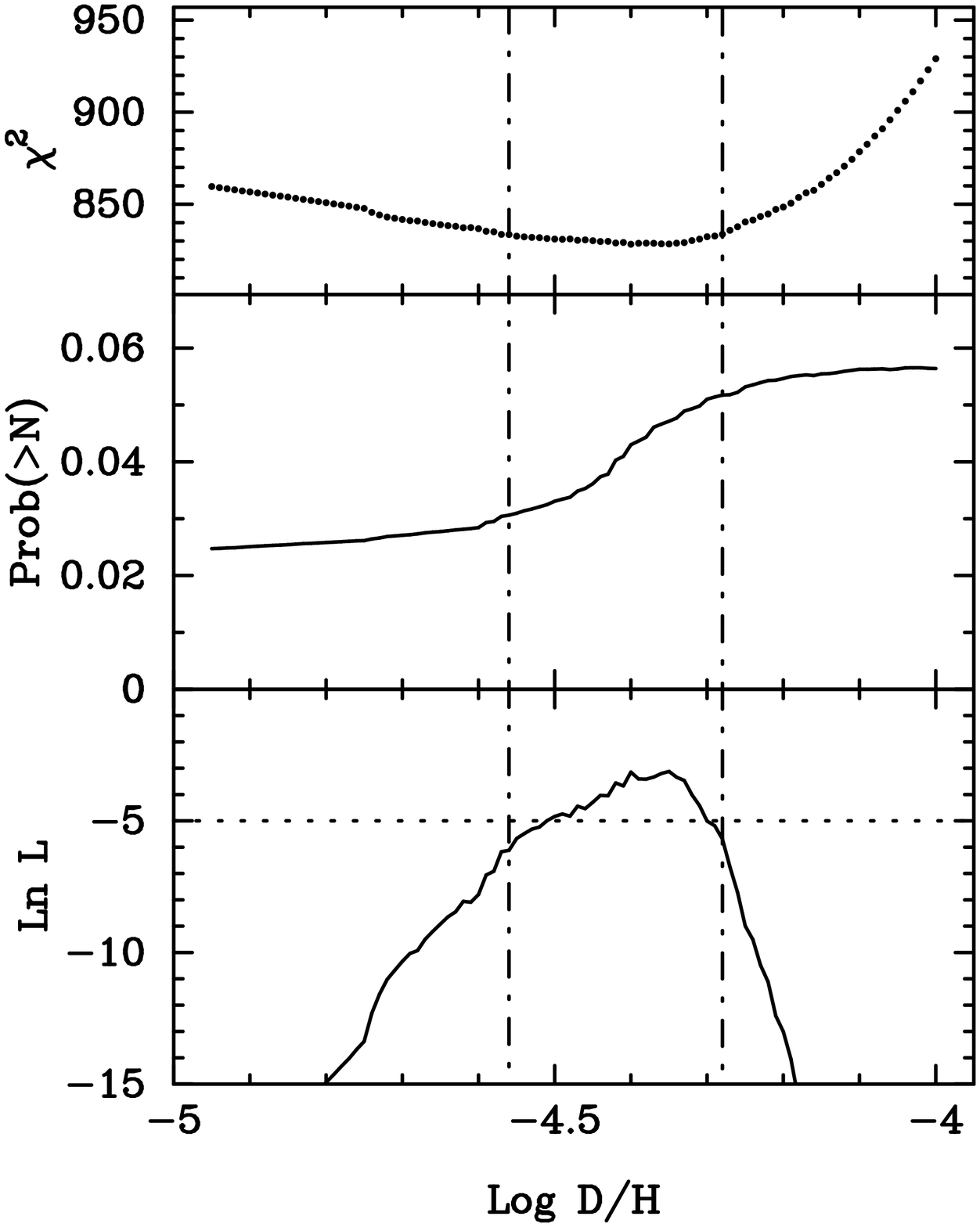,height=8.25 in,width=\columnwidth}}
\caption{
Likelihood of D/H with contamination.
The top panel shows the $\chi^2$ function from Fig. 9, the
middle panel shows the probability of a random \Lya line with a minimum
column density falling near deuterium, and the bottom panel shows the
combined likelihood. 
}
\end{figure}
\clearpage

\begin{figure}
\figurenum{12}
\centerline{
\psfig{file=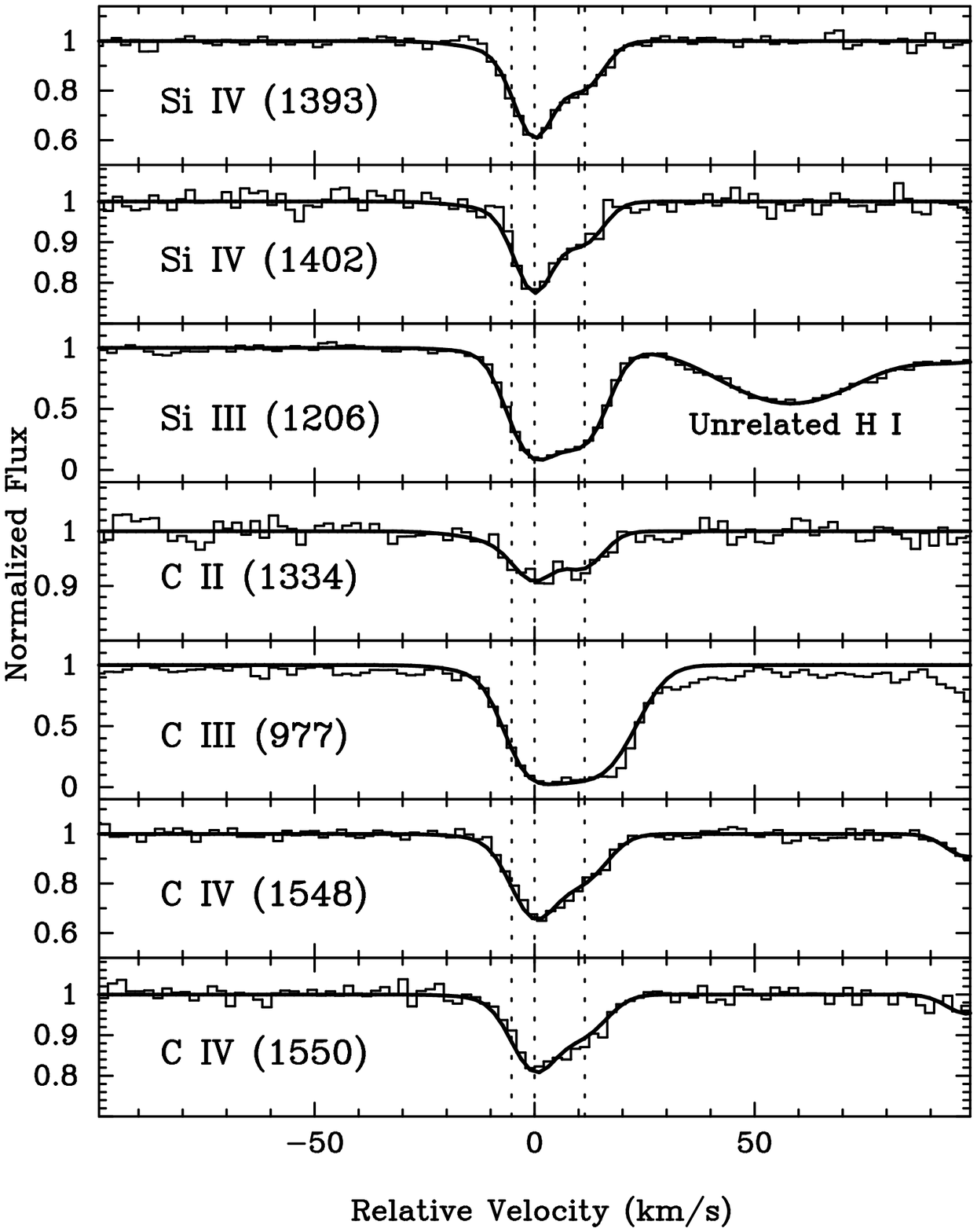,height=8.25 in,width=\columnwidth}}
\caption{
Zero velocity corresponds to redshift $z = 2.503571$ to match
Fig. 4.  The dotted lines correspond to absorption components
at the velocity positions given by the best fit to the
Lyman series in Model 6. 
}
\end{figure}
\clearpage

\begin{figure}
\figurenum{13}
\centerline{
\psfig{file=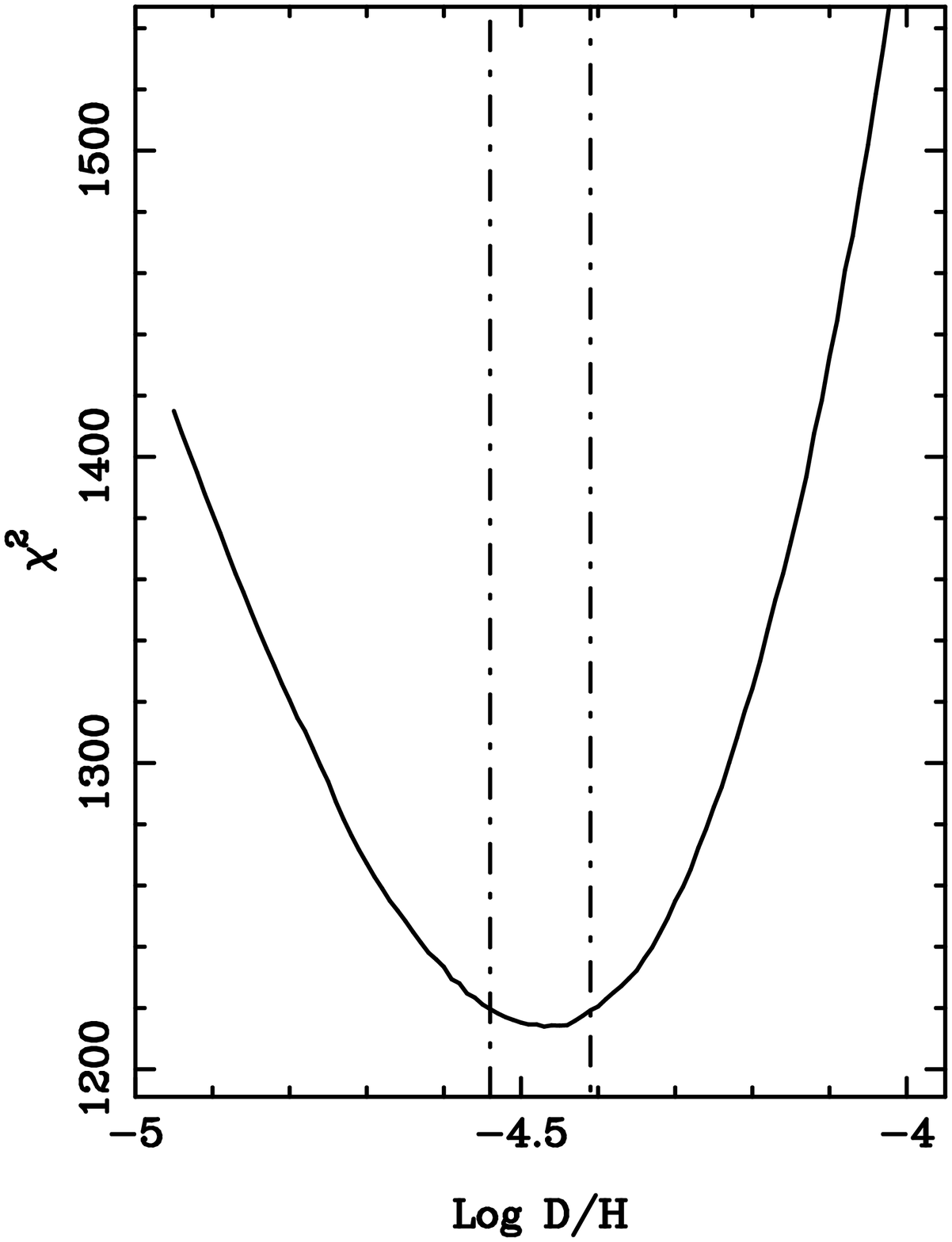,height=8.25 in,width=\columnwidth}}
\caption{
Total $\chi^2$ of Q1937--1009 and Q1009+2956 
}
\end{figure}
\clearpage

\begin{figure}
\figurenum{14}
\centerline{
\psfig{file=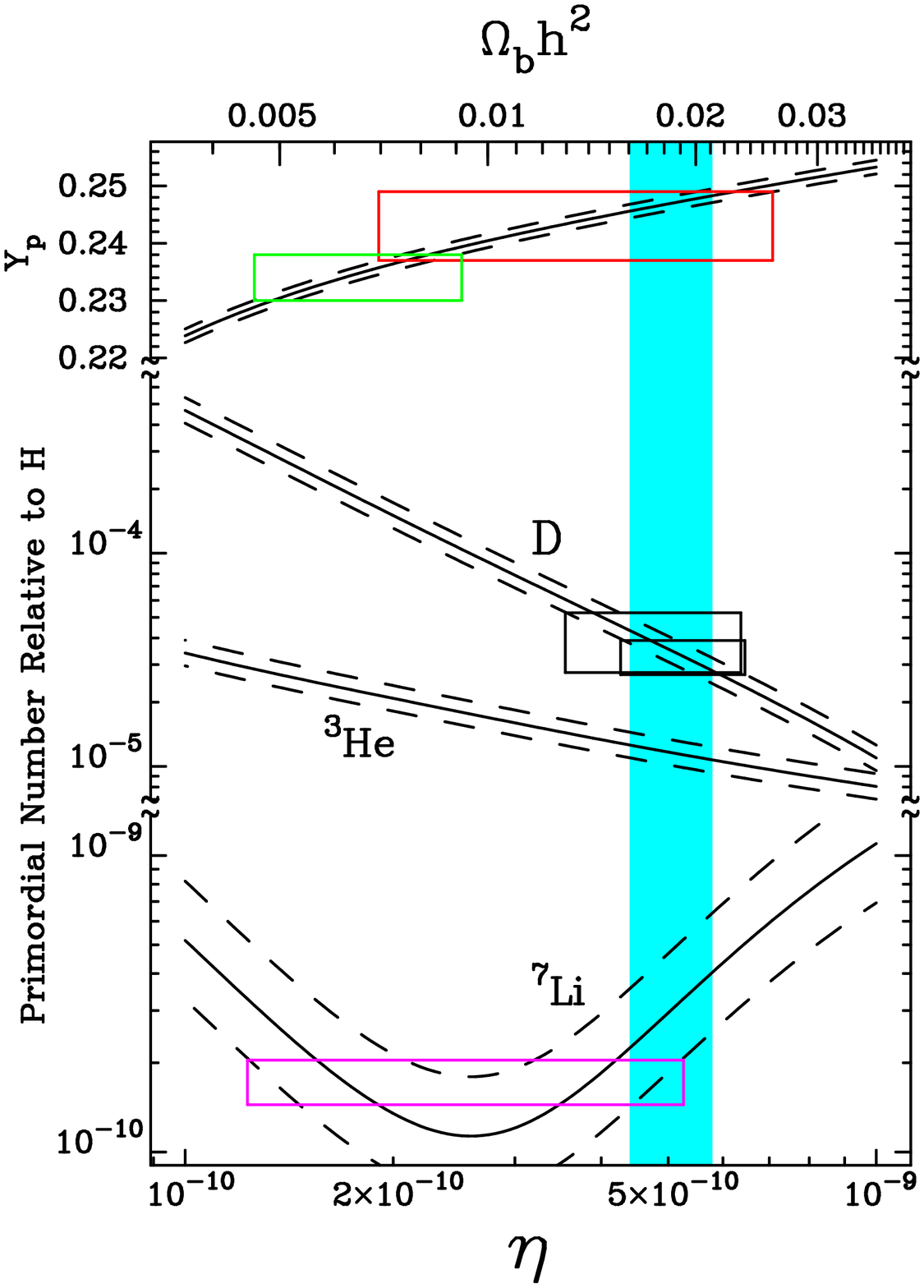,height=8.25 in,width=\columnwidth}}
\caption{
The predicted abundance ratios of the light elements
from SBBN as a function of $\eta$ and $\Omega_b \, h^2$.
$^4$He is shown as primordial mass fraction, Y$_p$.  Boxes represent
95\% confidence levels of recent observational determinations.
The width of the boxes include 95\% confidence levels
in the SBBN calculations.
}
\end{figure}

\end{document}